\documentclass[12pt,english,floatfix,superscriptaddress,aps,prd,preprint,showkeys,nofootinbib]{revtex4}
\usepackage{amsmath}
\usepackage{amssymb}
\usepackage{amsbsy}
\usepackage{amsfonts}
\usepackage{amsopn}
\usepackage{amstext}
\usepackage{graphicx}
\usepackage{amssymb}
\usepackage{amsfonts}
\usepackage{amsmath}
\usepackage{graphicx}
\usepackage[english]{babel}
\usepackage{color}
\usepackage{slashed}
\usepackage{esint}
\usepackage[dvips]{epsfig}
\usepackage[dvips]{graphicx}
\usepackage{float}
\usepackage{units}
\usepackage{textcomp}
\usepackage{caption}
\usepackage{multirow}
\usepackage{mathrsfs}
\usepackage{bm}
\usepackage{amsmath}
\usepackage{amssymb}
\usepackage{amsthm}
\usepackage{amsfonts}
\usepackage{mathtools}
\usepackage{color}

\usepackage{hyperref}
\usepackage{slashed}
\usepackage{hyperref}
\usepackage{graphicx}
\usepackage{amssymb}
\usepackage{amsmath}
\usepackage{color}

\usepackage[normalem]{ulem}

\newcommand{\be}{\begin{equation}}
\newcommand{\ee}{\end{equation}}
\newcommand{\ben}{\begin{eqnarray}}
\newcommand{\een}{\end{eqnarray}}

\begin{document}


\title{Braneworlds in $f(\mathbb{Q})$ gravity\\
}

\author{J. E. G. Silva}
\email{euclides.silva@ufca.edu.br}
\affiliation{Universidade Federal do Cariri (UFCA), Av. Tenente Raimundo Rocha, \\ Cidade Universit\'{a}ria, Juazeiro do Norte, Cear\'{a}, CEP 63048-080, Brasil}
\affiliation{Universidade Federal do Cear\'a (UFC), Departamento de F\'isica,\\ Campus do Pici, Fortaleza - CE, C.P. 6030, 60455-760 - Brazil.}

\author{R. V. Maluf}
\email{r.v.maluf@fisica.ufc.br}
\affiliation{Universidade Federal do Cear\'a (UFC), Departamento de F\'isica,\\ Campus do Pici, Fortaleza - CE, C.P. 6030, 60455-760 - Brazil.}
\author{Gonzalo J. Olmo}
\email{gonzalo.olmo@uv.es}
\affiliation{Departamento de Física Teórica and IFIC, Centro Mixto Universidad de Valencia—CSIC. Universidad
de Valencia, Burjassot-46100, Valencia, Spain.}
\affiliation{Universidade Federal do Cear\'a (UFC), Departamento de F\'isica,\\ Campus do Pici, Fortaleza - CE, C.P. 6030, 60455-760 - Brazil.}

\author{C. A. S. Almeida}
\email{carlosf@fisica.ufc.br}
\affiliation{Universidade Federal do Cear\'a (UFC), Departamento de F\'isica,\\ Campus do Pici, Fortaleza - CE, C.P. 6030, 60455-760 - Brazil.}

\begin{abstract}
We propose a braneworld scenario in a modified symmetric teleparallel gravitational theory, where the dynamics for the gravitational field is encoded in the nonmetricity tensor rather than in the curvature. Assuming a single real scalar field with a sine-Gordon self-interaction, the generalized quadratic nonmetricity invariant $\mathbb{Q}$ controls the brane width while keeping the shape of the energy density. By considering power corrections of the invariant $\mathbb{Q}$ in the gravitational Lagrangian, the sine-Gordon potential is modified exhibiting new barriers and false vacuum. As a result, the domain wall brane obtains an inner structure, and it undergoes a splitting process. In addition, we also propose a non-minimal coupling between a bulk fermion field and the nonmetricity invariant $\mathbb{Q}$. Such geometric coupling leads to a massless chiral fermion bound to the 3-brane and a stable tower of non-localized massive states. 
\end{abstract}


\maketitle

\section{introduction}

Despite the great success of general relativity (GR) in describing the gravitational effects in currently accessible weak and strong gravity astrophysical scenarios, open problems at larger scales and the lack of understanding of the nature of dark matter \cite{darkmatter} and dark energy \cite{darkenergy}  boost the interest in alternative theories of gravity. Among the many alternatives explored in the literature, one finds theories with new scalar and vector dynamical degrees of freedom \cite{Capozziello:2011et}, others based on massive gravitons \cite{Hinterbichler:2011tt},  string inspired constructions such as brane-worlds \cite{Maartens:2010ar}, and also geometric scenarios that break with the tradition set by Riemannian geometry. In this latter case, one can identify theories based on the Einstein-Cartan geometry \cite{einsteincartan}, metric-affine (or Palatini) theories of different kinds \cite{metricaffine}, including $f(R)$ models \cite{fr}, and also theories whose dynamics is entirely based on torsion and/or nonmetricity \cite{Arcos2004zh,fq3}. 

In theories such as the so-called teleparallel {equivalent} of general relativity (TEGR), gravity is understood as an effect of the spacetime torsion, rather than as a manifestation of curvature \cite{Aldrovandi}. In fact, the Riemann curvature tensor of the affine connection is assumed to vanish over the whole spacetime, whereas the torsion tensor is non-null \cite{Aldrovandi}. First proposed by Einstein, as an attempt to geometrize the electromagnetic field, the TEGR was latter recognized as a gauge theory of gravity based on the Poincar\'{e} group \cite{Baez:2012bn,Hohmann:2017duq,Maluf:2013gaa}. Instead of the metric tensor, the dynamical variable of the TEGR is the vielbein field. The vanishing curvature condition allows one to define a connection entirely in {terms} of the vielbeins, known as  Weitzenb\"{o}ck connection \cite{weit,Aldrovandi}. As a result, the same number of dynamical degrees of freedom of GR are also present in TEGR. Modifications of the framework set by TEGR, whose Lagrangian density is defined in terms of the torsion scalar $T$, have been studied in the last years in different forms, including theories of the $f(T)$ type \cite{ftinflation,Cai} and  $f(T,B)$ type \cite{ftb}, among others. In the latter ones, $B$ represents a boundary term in the TEGR case.

Conceptually related to the above, some interest has also grown recently in relation with what is  known as symmetric teleparallel equivalent of general relativity (STEGR), where the gravitational degrees of freedom are now described not in terms of the metric alone or the torsion alone, but by means of the nonmetricity tensor $Q_{\mu\nu\rho}=\nabla_\mu g_{\nu\rho}$ \cite{fq1}. Unlike TEGR, the dynamical field in STEGR does involve the metric tensor, hence the term symmetric. The gravitational action in STEGR is {built} using contractions of the metric, and the nonmetricity tensor up to quadratic order \cite{fq1,fq2}. For a specific choice of the coefficients of the quadratic terms, the resulting theory turns out to lead to the same equations of motion (EOM) as in GR. As expected, extensions of the basic STEGR framework were proposed by generalizing the gravitational action with some free coefficients, giving rise to what is known as symmetric teleparallel gravity (STG) \cite{fq3,delhom}, based on a scalar quantity denoted as $Q$,  or even considering $f(Q)$ modified Lagrangian terms \cite{fqenergyconditions}. STG has more propagating degrees of freedom than GR depending on the choice of the coefficients in the Lagrangian \cite{fqpropagation1,fqpropagation2}. The phenomenology of $f(Q)$ models has been studied in cosmology \cite{fqcosmology1,fqcosmology2}, black holes, \cite{fqblackhole} and wormhole \cite{fqwormhole} scenarios, among others. Our particular interest about $f(Q)$ gravity lies in the fact that this kind of non-linear extension of STEGR is not equivalent to $f(R)$ gravity, and both the metric and the affine connection carry physical information since the dependence on $\Gamma^{\lambda}_{\ \mu\nu}$ can no longer be absorbed in a boundary term \cite{fq1}.

Braneworld models have also offered interesting perspectives for high energy problems in the last decades. First proposed as a geometric solution for the gauge hierarchy problem, the warped geometry of the five-dimensional Randall-Sundrum (RS) model showed how the effective dynamics on a 3-brane can be altered by the dynamics of gravity in a higher dimensional spacetime (bulk) where the 3-brane is embedded in \cite{rs1}. In RS models the 3-brane is considered to be infinitely thin and the bulk is chosen to be an $AdS_5$ spacetime \cite{rs1,rs2}.
Soon after, models considering thick 3-branes as domain walls with one or more scalar fields as source were proposed \cite{domainwall}. 

On the other hand, in order to ensure brane stability, the so-called Bogomolnyi-Prasad-Sommerfield (BPS) solutions were found by employing a first-order formalism \cite{Gremm, wilami,blochbrane,compacton}. The propagation of gravity and bulk gauge and matter fields along the extra dimension can be analyzed by means of their Kaluza-Klein spectrum. Even for an infinite extra dimension, the bulk curvature allows for the existence of one gravitational massless normalized mode, reproducing the usual GR dynamics in the 3-brane. For the localization of the gauge fields, an additional dilaton coupling was considered, whereas for the matter (spinor) fields, a Yukawa-like coupling with a bulk scalar was proposed \cite{kehagias}.

Modified braneworld scenarios exploring wider geometric structures have also been considered, such as in Einstein-Cartan gravity \cite{roldao}, mimetic gravity \cite{minetic}, Weyl geometry \cite{weyl}, non-trivial inner manifold \cite{conifold,cigar}, with Palatini $f(R)$ dynamics \cite{fR1}, etc. In Ref.(\cite{Yang2012}), an $f(T)$ thick braneworld was investigated considering a non-quadratic Lagrangian. As a result, the thick brane exhibited an inner structure, even for a single scalar field. By varying the non-quadratic parameter, the brane undergoes a splitting transition \cite{Yang2012,Menezes,ftnoncanonicalscalar}. Similar results were found considering $f(T,B)$ dynamics and $f(T)$ in six dimensions \cite{allan1,Moreira:2021fva}. Recently, an STEGR $f(Q)$ braneworld has been proposed in Ref.\cite{fqbrane}, studying thick brane solutions and the propagation of gravitons in the extra dimension.

In this work, we propose a general STG codimension one braneworld scenario, whose quadratic invariant $\mathbb{Q}$ depends on several arbitrary coefficients and the gravitational Lagrangian is represented by a smooth (arbitrary) function $f(\mathbb{Q})$. We investigate the constraints imposed by the dimensional reduction and the regularity of the solutions on the coefficients of $\mathbb{Q}$, and also consider a function of the form $f(\mathbb{Q})=\mathbb{Q} + k \mathbb{Q}^n$, where $k$ and $n$ can be tuned to explore new phenomenology. For $n=2$, we find that the source scalar field exhibits a hybrid kink-compacton behavior, and the 3-brane undergoes a splitting process. In addition, we propose a non-minimal coupling between the nonmetricity scalar $\mathbb{Q}$ and a bulk fermion. We show that such coupling is equivalent to the commonly used Yukawa interaction. Furthermore, the effects produced by a given change in the $\mathbb{Q}$-coefficients and on the non-quadratic term involving the fermions are considered.

The work is organized as follows. In section \ref{sec2}, a brief review of the symmetric teleparallel gravity is presented, as well as its STEGR limit. In section \ref{sec3}, the $f(\mathbb{Q})$ braneworld is introduced, and the respective modified gravitational equations are studied. The vacuum, thin brane, and thick brane solutions are found and their properties are analysed. In section \ref{sec4}, we propose a geometric non-minimal coupling between the nonmetricity invariant $\mathbb{Q}$ and a bulk massless fermion and its effects are investigated. Final remarks are discussed in section \ref{sec5}.


\section{Symmetric teleparallelism} \label{sec2}

In this work we consider the symmetric teleparallel gravity theory in a metric-affine approach, whereby the metric $g_{MN}$ and the connection $\Gamma^{A}_{\ MN}$ represent independent degrees of freedom.

By parallel transporting the vector $V^{A}$, the 1-form covariant derivative takes the form $\nabla V^A = dv^{A}+\omega^{A}_{\ M}V^{M}$, where the connection 1-form is $\omega^{A}_{\ M}=\Gamma^{A}_{\phantom{A}NM}dx^N$. The covariant derivative of the metric $g_{MN}$ leads to the so-called nonmetricity 1-form
$\nabla g_{MN}=dg_{MN} -\omega^{A}_{\ M}g_{AN} - \omega^{A}_{\ N}g_{AM}$, where the nonmetricity tensor is defined by \cite{fq1}
\be
Q_{AMN}\equiv \nabla_{A}g_{MN}.
\ee
The torsion 2-form $T^{M}=\omega^{M}_{\ N}\wedge dx^N$ is another fundamental object associated to the affine connection, and its components define the torsion tensor ${T}^{A}_{\phantom{\alpha}BC} \equiv 2\Gamma^{A}_{\phantom{\alpha}[BC]}$
\cite{Aldrovandi,fq1}.
Thus, a general affine connection can be decomposed into a Levi-Civita part associated to the metric, and two more pieces defined in terms of the nonmetricity and the torsion tensors as \cite{fq1}
\be \label{decomposition}
\Gamma^{A}_{\phantom{\alpha}BC}=\left\{^{\phantom{i} A}_{BC}\right\}  + L^{A}_{\phantom{\alpha}BC}\, + K^{A}_{\phantom{\alpha}BC},
\ee with $\left\{^{\phantom{i} A}_{BC}\right\} \equiv \frac{1}{2}g^{AD}\left( \partial_{B}g_{DC} + \partial_{C}g_{DB}  - \partial_{D}g_{BC}\right)$ corresponding to the symmetric Levi-Civita connection $\mathcal{D}_{A}$, satisfying $\mathcal{D}_{A} g_{BC}=0$. The second (symmetric) term  $L^{A}_{\ BC}$, called the disformation tensor, is completely determined by the nonmetricity tensor according to 
\be \label{disformation}
L^{A}_{\phantom{\alpha}BC}  \equiv \frac{1}{2} Q^{A}_{\phantom{\alpha}BC} - Q_{(B\phantom{\alpha}C)}^{\phantom{(\mu}A}\,. 
\ee  
The antisymmetric piece $K^{A}_{\ BC}$ is known as contortion tensor and it is solely determined by the torsion tensor as
\be \label{contortion}
K^{A}_{\phantom{\alpha}BC} \equiv \frac{1}{2}T^{A}_{\phantom{\alpha}BC} + T_{(B{\phantom{\alpha}C)}}^{\phantom{,\mu}A}\,.
\ee
In symmetric teleparallel theories, torsion is assumed to vanish everywhere. For this reason, from now on we only consider torsion free connections.

From the nonmetricity tensor it is possible to define the trace vectors as \cite{fq1}
\be \label{qtrace}
Q^{A} =g^{MN} Q^{A}_{\ M N}\,, \quad \bar{Q}_{A}=\bar{Q}^{H}_{\phantom{\alpha}AH}\,,
\ee
where the traceless part has been separated as
\be
\bar{Q}_{ABC}  =  Q_{ABC} -  \frac{1}{4}g_{BC}Q_{A}\,.
\ee
Moreover, it is convenient to introduce the so-called superpotential tensor defined as \cite{fq1,fq2,fq3}
\be \label{nmsuperquad}
P^{K}{}_{MN} = c_1 Q^{K}{}_{MN}+c_2Q_{(M\phantom{\alpha}N)}^{\ \ \phantom{\mu}K}+c_3 Q^{K} g_{MN} +  c_4\delta^{K}_{(M}\bar{Q}_{N)} + \frac{c_5}{2}\big(\bar{Q}^{K} g_{MN}+\delta^{K}_{(M}Q_{N)}\big) \, ,
\ee 
{where the five coefficients $c_{i}$ are, for the moment, arbitrary real constants.}
The contraction of the $P$ and $Q$ tensors provides a quadratic invariant called the generalized nonmetricity scalar by
\be
\mathbb{Q}=Q_{T}^{\  MN}P^{T}_{\ MN}.
\ee
{For a specific choice of coefficients given by} \cite{fq1}
\be \label{qgr}
c_1 = -c_3= -\frac{1}{4}\,, \quad 
c_2 = -c_5= \frac{1}{2}\,, \quad
c_4 = 0\,,
\ee
the nonmetricity scalar leads to
\be \label{q2}
\mathbb{Q}=-\mathcal{Q} =   \frac{1}{4}Q_{KMN}Q^{KMN} -  \frac{1}{2}Q_{KMN}Q^{MNK} 
  -   \frac{1}{4}Q_{K} Q^{K}  
  + \frac{1}{2}Q_{K}\bar{Q}^{K}.
\ee
With the above notation, the Ricci scalar for the connection $\Gamma^{T}_{MN}$ {can be written in the form}
\be \label{ricciscalarq}
R   =  \mathcal{R}  + \mathcal{Q} +   \mathcal{D}_{A}( Q^{A} - \bar{Q}^{A} )\,.
\ee
In teleparallel theories, the Riemann tensor defined by the affine connection $\Gamma^{T}_{MN}$ vanishes by construction and so does its associated Ricci scalar $R$. Thus, the Ricci scalar corresponding to the Levi-Civita connection $\mathcal{R}$ is related to the nonmetricity scalar by
\be \label{ricciscalarq}
\mathcal{R}= -\mathcal{Q} -\mathcal{D}_{A}( Q^{A} - \bar{Q}^{A} )\,.
\ee
Therefore, the Einstein-Hilbert Lagrangian $\mathcal{L}_{EH}=\frac{1}{2\kappa}\sqrt{-g} \mathcal{R}$ is equivalente to the symmetric teleparallel Lagrangian $\mathcal{L}_{\mathcal{Q}}=-\frac{1}{2\kappa}\sqrt{-g} \mathcal{Q}$. For this obvious reason the choice of parameters (\ref{qgr}) is called the {\it symmetric teleparallel equivalent of general relativity} (STEGR) \cite{fq1}.

Another consequence of the vanishing Riemann tensor for the connection $\Gamma^{T}_{MN}$ is that we can choose a class of connections satisfying this condition. A particular choice, known as the \textit{coincident gauge}, is achieved by simply assuming that $\Gamma^{T}_{MN}=0$. As a result, $L^{T}_{MN}=-\left\{^{\phantom{i} T}_{MN}\right\} $ \cite{fq1,fq2,fqcosmology1}.

In this work we are interested in modified gravitational theories inspired by the STEGR and, for this reason, we consider an action of the form
\begin{eqnarray}
\label{action1}
S&=&\int{d^{D}x \sqrt{-g}\left(\frac{1}{2\kappa_D} f(\mathbb{Q}) -2\Lambda +\mathcal{L}_{M}\right)},
\end{eqnarray}
where {$\kappa_{D}=8\pi G_{D}$ with $G_{D}$ being the $D$-dimensional} Newtonian gravitational constant, $\Lambda$ represents the bulk cosmological constant, $f(\mathbb{Q})$ is a smooth function of the nonmetricity scalar $\mathbb{Q}$, and $\mathcal{L}_{m}$ represents the matter Lagrangian density. Variation of the action (\ref{action1}) with respect to the metric leads to the gravitational field equations (where $f_\mathbb{Q}\equiv \frac{d f}{d \mathbb{Q}}$)
\begin{eqnarray}
\label{fQeom}
\frac{2}{\sqrt{-g}}\nabla_{K}\left(\sqrt{-g}f_{\mathbb{Q}} P^{K}_{\ MN}\right)-\frac{\left(f-2\Lambda\right)}{2} g_{MN}+f_\mathbb{Q}\left(P_{MKL}Q_{N}^{\ KL}-2Q_{KM}^{\ \ \ L}P^{K}_{\ NL} \right)= \kappa \mathfrak{T}_{MN}.
\end{eqnarray} 
The energy-momentum tensor is given by 
\be
\mathfrak{T}_{MN} = -\frac{2}{\sqrt{-g}}\frac{\delta \left( \sqrt{-g}\mathcal{L}_{m}\right) }{\delta g^{MN}} \ .
\ee 
In the following we will use the expressions of above in scenarios involving braneworlds. 


\section{Symmetric teleparallel braneworld}
\label{sec3}

Consider a warped five-dimensional spacetime represented by the
following line element
\be\label{RSmetric}
ds^2{}_5=e^{2A(y)}\hat{g}_{\mu\nu}(x)dx^{\mu}dx^{\nu}+e^{2B(y)}dy^{2},
\ee
where $A(y)$ and $B(y)$ are warp factors that depend only on the extra-dimensional coordinate $y$. The Greek indices $\mu$, $\nu$, $\dots$ run from 0 to 3. The 3-brane metric $\hat{g}_{\mu\nu}$ depends only on the brane coordinates $x^\mu =(x^0 , \Vec{r})$ and for a flat 3-brane, $\hat{g}_{\mu\nu}=\eta_{\mu\nu}$.  By assuming a non-flat brane $\hat{g}_{\mu\nu}$, gravitational effects along the brane can be considered. 

Let us now compute the relevant objects for the ansatz (\ref{RSmetric}) and in the coincident gauge, with $\Gamma^{A}_{BC}=0$, so that the covariant derivatives are simply partial derivatives. The only non-vanishing components of the nonmetricity tensor are 
\be \label{Qfrw}
Q_{4 \alpha\beta} = 2A'e^{2A}\hat{g}_{\alpha\beta}\,, \quad Q_{444} = 2B'e^{2B} \,, \quad \hat{Q}_{\rho\mu\nu}=\partial_\rho \hat{g}_{\mu\nu},
\ee where prime stands for derivative with respect to $y$. Henceforth we will assume a Gaussian configuration to generate a thick braneworld where $B=0$, then  the 5-D nonmetricity scalar $\mathbb{Q}$ is given by
\begin{equation}
    \mathbb{Q}=e^{-2A}\hat{\mathbb{Q}} + 16 (c_1 + 4c_3)A'^2,
\end{equation} where the induced nonmetricity scalar $\hat{\mathbb{Q}}$ is made up with the four-dimensional metric and has no dependence on the extra-dimensional coordinate $y$. Then, one may consider the quadratic action
\begin{eqnarray}
\label{action2}
S&=&\frac{1}{2\kappa_{5}}\int{d^{5}x \sqrt{-g}\mathbb{Q}},
\end{eqnarray}
and integrating out on the extra dimension we can write the 4-D effective gravitational action in the form
\begin{eqnarray}
\label{action2}
S_4&=&\frac{1}{2\kappa_4}\int{d^{4}x \sqrt{-\hat{g}}\mathbb{\hat{\mathbb{Q}}}},
\end{eqnarray}
where we assume the condition $A'(0)=0$ and define 
\begin{eqnarray}
\kappa_4 \equiv \frac{k_5}{\int_{-\infty}^{\infty}e^{2A}dy}.
\end{eqnarray}
Thus, the dimensional reduction leads to a finite 4-D effective gravitational constant $\kappa_4$ provided that $\int_{-\infty}^{\infty}e^{2A}dy$ is finite. For a thin brane in $AdS_5$ (RS model), $A(y)=-c |y|$ and thus, $\kappa_4 = c k_5$. Since $A(y)$ is a solution of the gravitational equations, $\kappa_4$ should depend on the coefficients $c_i$ of the gravity Lagrangian.
Indeed, in the thin limit of the modified quadratic field equations we obtained this relation.

Now consider a modified gravitational theory such as
\begin{equation}
    S=\frac{1}{2\kappa_{5}}\int{d^{5}x \sqrt{-g}f(\mathbb{Q})},
\end{equation}where we consider the case of particular interest $f(\mathbb{Q})=\mathbb{Q}+k \mathbb{Q}^{n}$ with $k$ being a constant with mass dimension $[k]=M^{2-2n}$, and $n$ is a real number. If $\hat{\mathbb{Q}}\neq 0$, then the power-law term leads to an effective 4-D action
\begin{equation}
S_4 =\frac{k}{2\kappa_{4}}\int{d^{4}x \sqrt{-\hat{g}}e^{(4-2n)A} \hat{\mathbb{Q}}^n}
\end{equation}
which is finite only for $n< 2$. Thus, an effective 4-dimensional gravitational action can only be well-defined for $n<2$. Therefore, the warped geometry constrains the possible modified gravitational actions for a bent 3-brane. That is a noteworthy constrains to be considered in cosmological branes or to recognize the gravitational dynamics along the brane as an effective dynamics from a higher dimensional spacetime.

\subsection{Flat brane with a bulk scalar field}
We will now focus on the gravitational effects along the extra dimension. For this purpose, we consider a flat 3-brane whose source is a minimally coupled real scalar field in the form
\be
\mathcal{S}_M=\int\boldsymbol{{\rm d}}^5x\sqrt{-g}\left( -\frac{1}{2}g^{AB}\nabla_{A}\phi\nabla_{B}\phi-V(\phi)\right),\label{Smatter}
\ee where $\phi\equiv \phi(y)$ depends only on the extra dimension $y$. The potential $V$ provides domain-wall solutions which guarantee the stability of the 3-brane.

The equations of motion for the metric and the scalar field from actions (\ref{action1}) and (\ref{Smatter}) take the form,
\begin{eqnarray}
4(c_{1}+4c_{3})\left[\left(A''+4A'^{2}\right)f_{\mathbb{Q}}+A'f'_{\mathbb{Q}}\right]  - \frac{1}{2}f+\kappa_5 V+\frac{1}{2}\kappa_5\phi'^{2} + \Lambda	& = & 0 \label{braneequation1}\\ 
8(c_{5}+2c_{3})\left(A''f_{\mathbb{Q}}+A'f'_{\mathbb{Q}}\right)+16(c_{1}+2c_{5}+8c_{3})A'^{2}f_{\mathbb{Q}}-\frac{1}{2}f+\kappa_5 V-\frac{1}{2}\kappa_5 \phi'^{2}+\Lambda & = & 0 \label{braneequation2}\\
\phi'' + 4 \phi' A'-V_{\phi}	& = & 0 \label{braneequation3} \ ,
\end{eqnarray}
where we use the metric ansatz  \eqref{RSmetric} together with the conditions $\hat{g}_{\mu\nu}=\eta_{\mu\nu}$ and $B=0$. It is worth noting that in the particular case in which we recover GR, when $c_1 = -c_3= -1/4,\ c_2 = -c_5= 1/2,\ c_4 = 0$, the above equations coincide with those obtained in\footnote{We also mention that our analysis here departs from that of Ref. \cite{fqbrane} regarding scalar fields in that we also consider effects due to a generalized superpotential (\ref{nmsuperquad}).} Ref. \cite{fqbrane}. Moreover, it is easy to see that there are only two free effective parameters in those equations, namely $\sigma_0=c_{5}+2c_{3}$ and $\sigma_1\equiv c_{1}+4c_{3}$, whose values in the STEGR case are just $\sigma_0^{GR}= 0$ and $\sigma_1^{GR}=3/4$. Thus, there are still three parameters in the theory that do not influence the background brane solution and whose impact in the physics should be explored by other means.

The modified gravitational equations can be recast into a more familiar form as
\begin{eqnarray}
\label{eom1}
    6A'^2&=&-\frac{\kappa_5}{f_{\mathbb{Q}}}\left(p_4 + \frac{f_{\mathbb{Q}}}{\kappa_5}\tilde{p}_4\right),
\end{eqnarray}
 \begin{eqnarray}
 3A'' + 6A'^2&=&-\frac{\kappa_5}{f_{\mathbb{Q}}}\left(\rho + \frac{f_{\mathbb{Q}}}{\kappa_5}\tilde{\rho}\right),   
\end{eqnarray}
where we encompassed all the modifications in the functions $\tilde{p}$ and $\tilde{\rho}$ given by
\begin{eqnarray}
    \tilde{p}_4 &=& 8\sigma_0A'' + [6+32\sigma_0]A'^2 + 8\sigma_0\frac{f'_{\mathbb{Q}}}{f_{\mathbb{Q}}} -\frac{f}{2f_{\mathbb{Q}}}\\
    \tilde{\rho}&=&\left(4\sigma_1-3\right)A'' +12\left(2\sigma_1-1\right)A'^2+4\sigma_1\frac{f'_{\mathbb{Q}}}{f_{\mathbb{Q}}} -\frac{f}{2f_{\mathbb{Q}}}.
    \label{geoenergy}
\end{eqnarray}
Note that $\tilde p$ only depends on $\sigma_0$, while $\tilde\rho$ only on $\sigma_1$. Thus,  $f(\mathbb{Q})$ theories modify the gravitational equations introducing geometric energy density $\tilde{\rho}$ and pressure $\tilde{p}_4$, though the degree of differentiability of the equations does not change as compared to STEGR. In the absence of the terms $\tilde{\rho}$ and $\tilde{p}_4$, we recover the GR-based braneworld Einstein equations \cite{domainwall,Gremm,wilami}.

Let us analyze the configurations such that $\tilde{\rho}\geq 0$. For the second term in (\ref{geoenergy}) to be non-negative, we impose that
\begin{equation}
\sigma_1 \geq 1/2.
\end{equation}  
In a warped compactified brane one expects that $e^{A}\rightarrow 0$, as $y \rightarrow \pm \infty$. Thus, assuming the conditions
\begin{equation}
   A'(0)= 0  \,  , A''(0)<0,
\end{equation}
the first term is non-negative near the origin (brane core) provided that
\begin{equation}
    \sigma_1 \leq  3/4.
\end{equation}
We recall that $\sigma_1^{GR}=3/4$, which is the case that sets the upper bound in  this inequality.
This condition restricts the range of $\sigma_1$ to the interval $ 1/2 \leq \sigma_1 \leq 3/4$.

As mentioned earlier, we are interested in a deviation of the STEGR in the form
\begin{eqnarray}
f(\mathbb{Q})=\mathbb{Q}+k\mathbb{Q}^{n}.
\end{eqnarray}
Since $\mathbb{Q}=16\sigma_1A'^{2}$, the effective gravitational action on the brane vanishes. Moreover, the third term in Eq.\eqref{geoenergy} is non-negative for  
\begin{equation}
\sigma_1 k \leq 0 \ .
\end{equation}
The last term in Eq.\eqref{geoenergy} vanishes at the origin and asymptotically. For $k<0$, the condition $f_Q >0$ leads to an upper limit for $A'$ given by
\begin{equation}
    A'<\left(\frac{1}{n|k|}\right)^{1/2(n-1)}.
\end{equation}
Therefore, assuming a weak energy condition for $\tilde{\rho}$, we have been able to put some constraints on the possible values of the coefficients $\sigma_1$ and $k$. For the pressure-like terms $\tilde{p}$, on the other hand, we obtain different equations of state for different values of the parameters, which may lead to inner structure effects, as we will see in the following sections.


\subsection{Quadratic gravity ($k=0$)}

In this regime, the gravitational Lagrangian is just determined by the generalized quadratic invariant $\mathbb{Q}$, and the field equations turn into
\be
\label{quadraticequaion1}
4\sigma_1A''+8\sigma_1A'^{2}+\kappa_5 V(\phi)+\frac{1}{2}\kappa_5 \phi'^{2}=- \Lambda,
\ee
\be 
\label{quadraticequation2}
8\sigma_0A''+8\left(\sigma_1+4\sigma_0\right)A'^{2}+\kappa_5 V(\phi)-\frac{1}{2}\kappa_5\phi'^{2}=- \Lambda.
\ee
As expected, the conditions (\ref{qgr}) lead to the well-known  braneworld gravitational equations of GR \cite{Gremm,wilami}.

\subsubsection{Outside the core}
Outside the brane core, where $\phi'=0=V(\phi)$, the gravitational equations \eqref{quadraticequaion1} and \eqref{quadraticequation2} boil down to 
\be
(\sigma_1-2 \sigma_0) A''(y)-8\sigma_0 A'(y)^2=0,
\ee 
which can be readily solved resulting in
\be\label{eq:bulkRS}
A(y)=C_1-\frac{(\sigma_1-2 \sigma_0)\ln\left[8 \sigma_0 |y| + C_2 (\sigma_1-2 \sigma_0)\right]}{8 \sigma_0},
\ee 
with $C_1$ and $C_2$ arbitrary integration constants. The limiting case $\sigma_0=0$ must be treated separately, yielding $A(y)=\tilde{C}_1+\tilde{C}_2|y|$, which is independent of $\sigma_1$. Since regularity at infinity demands a decaying $e^{2A}$ as $|y|\to \infty$, we conclude that for $\sigma_0=0$ we must have $\tilde{C}_2<0$. For $\sigma_0>0$, the dominant contribution is given by
\begin{equation}
\label{quadraticasymptotic}
    e^{2A}\approx \left[8\sigma_0 |y|\right]^{-\frac{(\sigma_1-2 \sigma_0)}{4\sigma_0}} ,
\end{equation}
which guarantees a rapid decay for any solution near the GR point ($\sigma_{1}\to 3/4, \sigma_0\to 0$). Interestingly, for negative $\sigma_0$ the bulk is only defined on the interval $0\leq |y|< C_2 (\sigma_1+2 |\sigma_0|)/8 |\sigma_0|$. Thus, $\sigma_0 <0$ leads to a compact extra dimension whose radius depends on the nonmetricity coefficients $c_1$,$c_3$ and $c_5$.


It is worth mentioning that, even though the solution (\ref{eq:bulkRS}) leads to $\mathbb{Q}=\frac{16\sigma_1(\sigma_1-2\sigma_2)}{y^2}$, which is singular at the origin, the solution in Eq. (\ref{quadraticasymptotic}) only represents the warp function outside the brane core and, therefore, the limit $y\to 0$ should be computed using the appropriate solution. 

Let us now consider a vacuum solution with bulk cosmological constant $\Lambda$, where the scalar field and the potential vanish, i.e., $\phi=V=0$. Assuming a RS solution as $A'=-c$ \cite{rs1},  the $\mathbb{Q}$-gravity equations lead to

\begin{equation}
\label{qgravitythinlimit}
    c^{2}=-\frac{1}{8}\frac{\Lambda}{(\sigma_{1} + 2 \sigma_0)} ,
\end{equation}
which relates the brane tension $c$ to the bulk cosmological constant $\Lambda$. As long as $(\sigma_{1} + 2 \sigma_0)>0$, one obtains the usual $AdS_5$ warped compactified bulk of the RS model, whereas for $(\sigma_{1} + 2 \sigma_0)<0$ it describes a $dS_5$ bulk. An obvious lesson that follows from this is that there exists a broad region of parameter space around the GR solution ($\sigma_{1}\to 3/4, \sigma_0\to 0$) for which the usual $AdS_5$ RS bulk is recovered, thus showing that such a solution is quite robust for not necessarily small departures from GR.  
The nonmetricity scalar for this solution is given by $\mathbb{Q}=-\frac{2\Lambda\sigma_1}{\sigma_1+2\sigma_0}$, which is well-behaved everywhere. This result also indicates that the solution (\ref{qgravitythinlimit}) can be regarded as a robust RS thin brane configuration.

\subsubsection{BPS solutions}
Now that we have gained some insight on the modifications induced by the $\mathbb{Q}$-gravity on the vacuum, it is time to turn our attention to the changes on the brane core and on the source (scalar field). This can be accomplished by means of the first-order formalism, wherein we seek for a BPS solution of the equations Eq.(\ref{braneequation3}), Eq.(\ref{quadraticequaion1}) and Eq.(\ref{quadraticequation2}) without bulk cosmological constant $(\Lambda=0)$. 

Consider a potential of the form
\begin{equation}
\label{quadraticscalarpotential}
    V(\Phi)=\alpha \left(\frac{\partial W}{\partial \Phi}\right)^2 -\frac{1}{3} W^2,
\end{equation}
together with the {first}-order system 
\begin{equation}
\label{quadraticbps1}
\frac{dA}{dy}=-\frac{1}{3} W 
\end{equation}

\begin{equation}
\label{quadraticbps2}
\frac{d\Phi}{dy}= \sqrt{2\alpha} \frac{\partial W}{\partial \Phi}.
\end{equation}
The BPS equations above satisfies the scalar field and the modified gravitational EOM provided that
\begin{equation}
    \alpha=\frac{\sigma_1 +  2\sigma_0}{6}.
\end{equation}
For the usual GR equivalent case, the constant $\alpha$ takes the value $\alpha=1/8$ \cite{Gremm,wilami}.

\begin{figure}[htb] 
       \begin{minipage}[b]{0.48 \linewidth}
           \includegraphics[width=\linewidth]{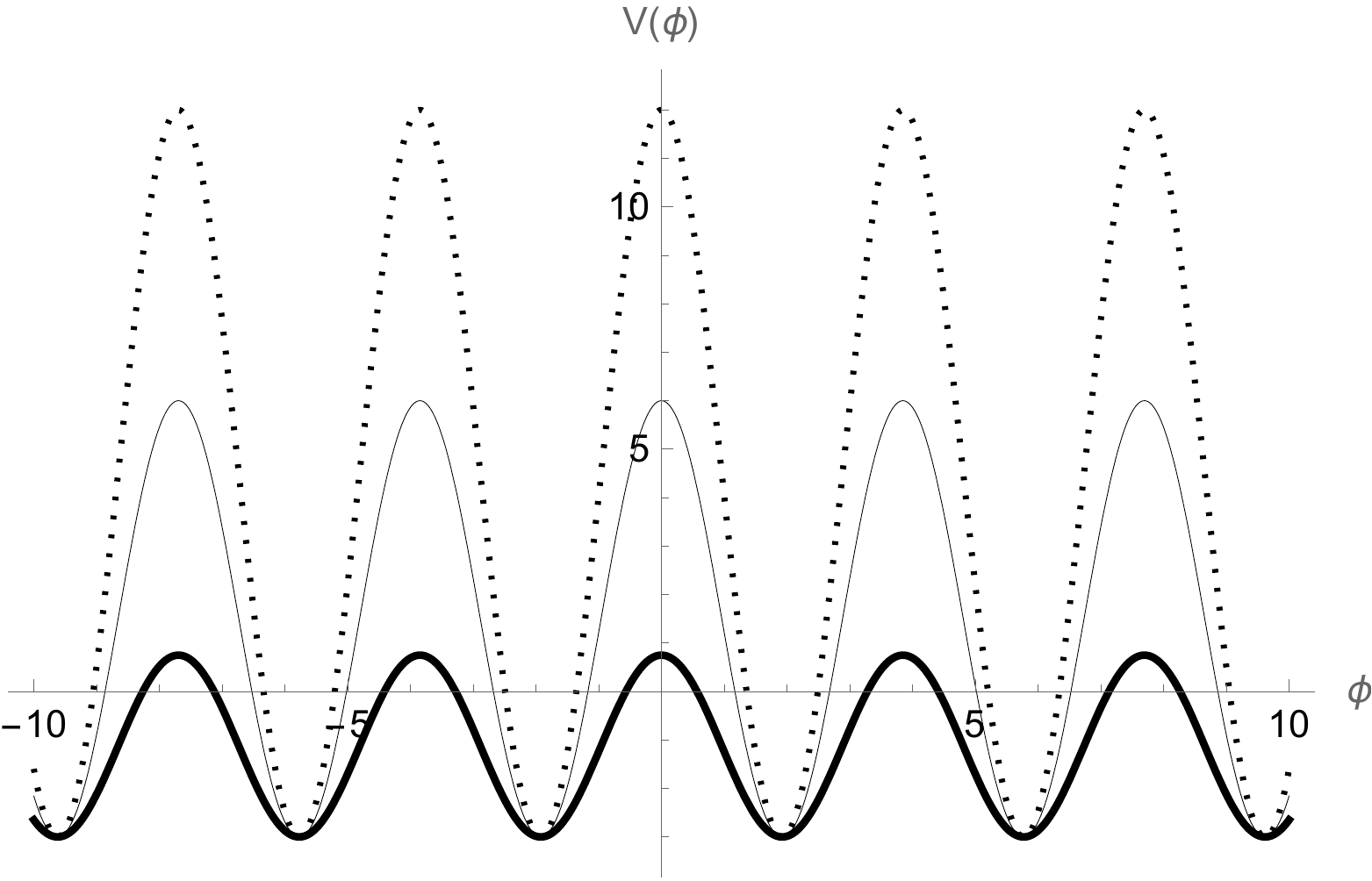}\\
           \caption{Modified sine-Gordon potential for $b=c=1$. For $\alpha=1/8$ (thick line), $\alpha=1$ (thin line) and $\alpha=2$ (dotted line) the vacuum points are preserved.}
          \label{quadracticscalarfieldpotential}
       \end{minipage}\hfill
       \begin{minipage}[b]{0.48 \linewidth}
           \includegraphics[width=\linewidth]{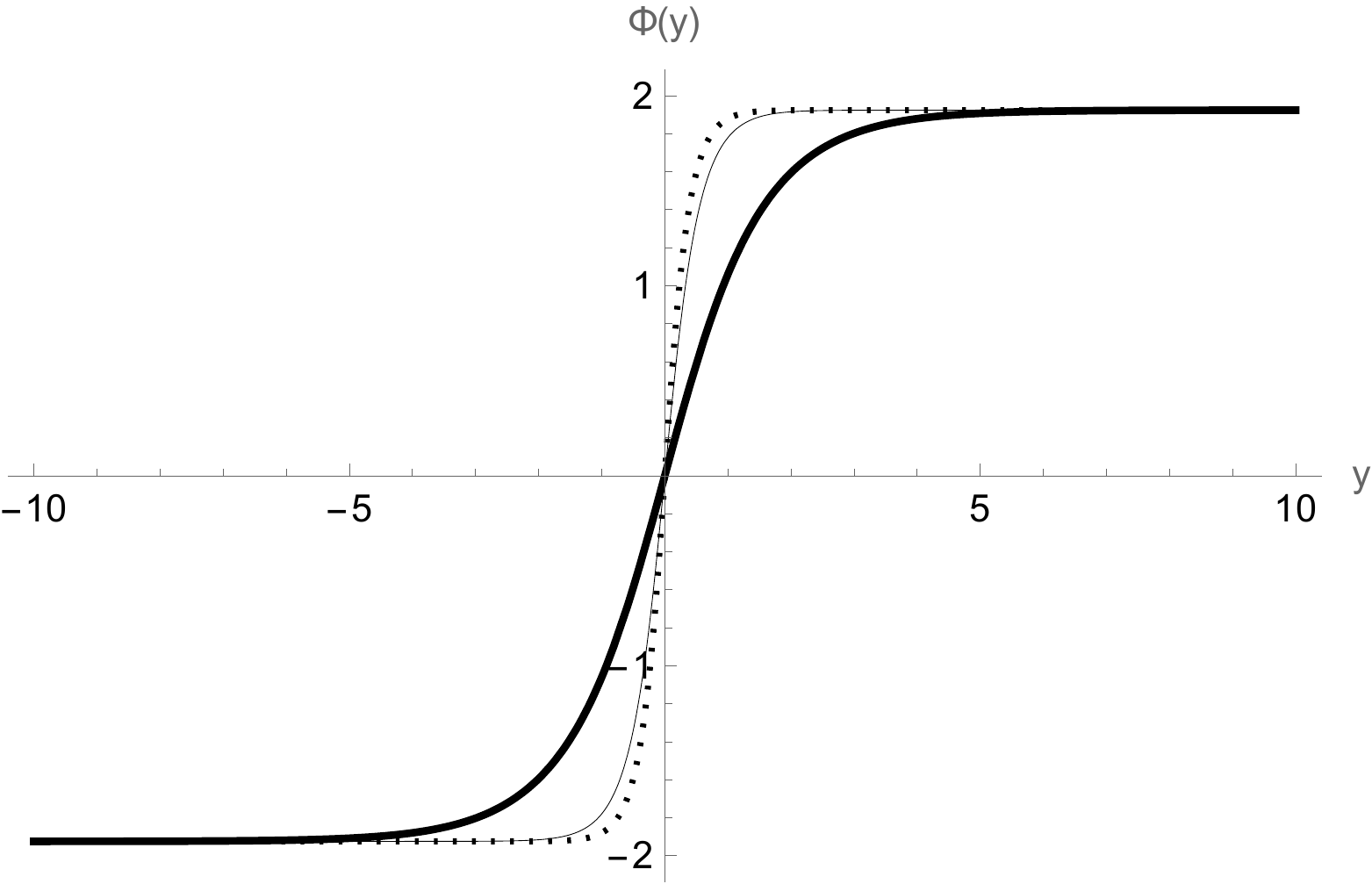}\\
           \caption{Scalar field profile for $b=c=1$. As $\alpha$ increases from $\alpha=1/8$ (thick line) and $\alpha=1$ (thin line) to $\alpha=2$ (dotted line) the field concentrates around the origin.}
           \label{quadraticscalarfieldprofile}
       \end{minipage}
   \end{figure}

Let us consider the well-known sine-{Gordon} model, where the superpotential is given by
\begin{equation}
    W(\phi)=3bc \sin{\Bigg[\sqrt{\frac{2}{3b}}\phi\Bigg]}.
\end{equation}
Using Eq.(\ref{quadraticscalarpotential}), the STG coefficients modified the sine-Gordon in the form
\begin{equation}
    V(\phi)=\frac{3bc^2}{2}\left((2\alpha -b) + (2\alpha+b)\cos{\left(2\sqrt{\frac{2}{3b}}\phi\right)}\right).
\end{equation}
A similar modified sine-Gordon potential was analyzed in Ref.(\cite{koley}).
As seen in fig.\ref{quadracticscalarfieldpotential}, the vaccua structure of the sine-Gordon potential is kept invariant as we vary the parameter $\alpha$. The scalar field solution of the BPS equation (\ref{quadraticbps2}) leads to
\begin{equation}
    \phi(y)=\sqrt{6b}\arctan(\tanh(\sqrt{2\alpha}cy)).
\end{equation}
Note that the higher the value of $\alpha$ the faster the scalar field attains the vacuum, as shown in Fig.(\ref{quadraticscalarfieldprofile}).  
By solving Eq.\ref{quadraticbps1} we obtain the warped function
\begin{equation}
\label{quadraticwarpedfunction}
    A(y)=\ln(sech(\lambda y))^{b/\sqrt{8\alpha}},
\end{equation}
where $\lambda=\sqrt{2\alpha}c$. Thus, the width of the thick brane is controlled by the nonmetricity coefficients $c_1$, $c_3$ and $c_5$ through their combination in the effective parameter $\alpha$.  

\begin{figure}
    \centering
    \includegraphics[scale=0.5]{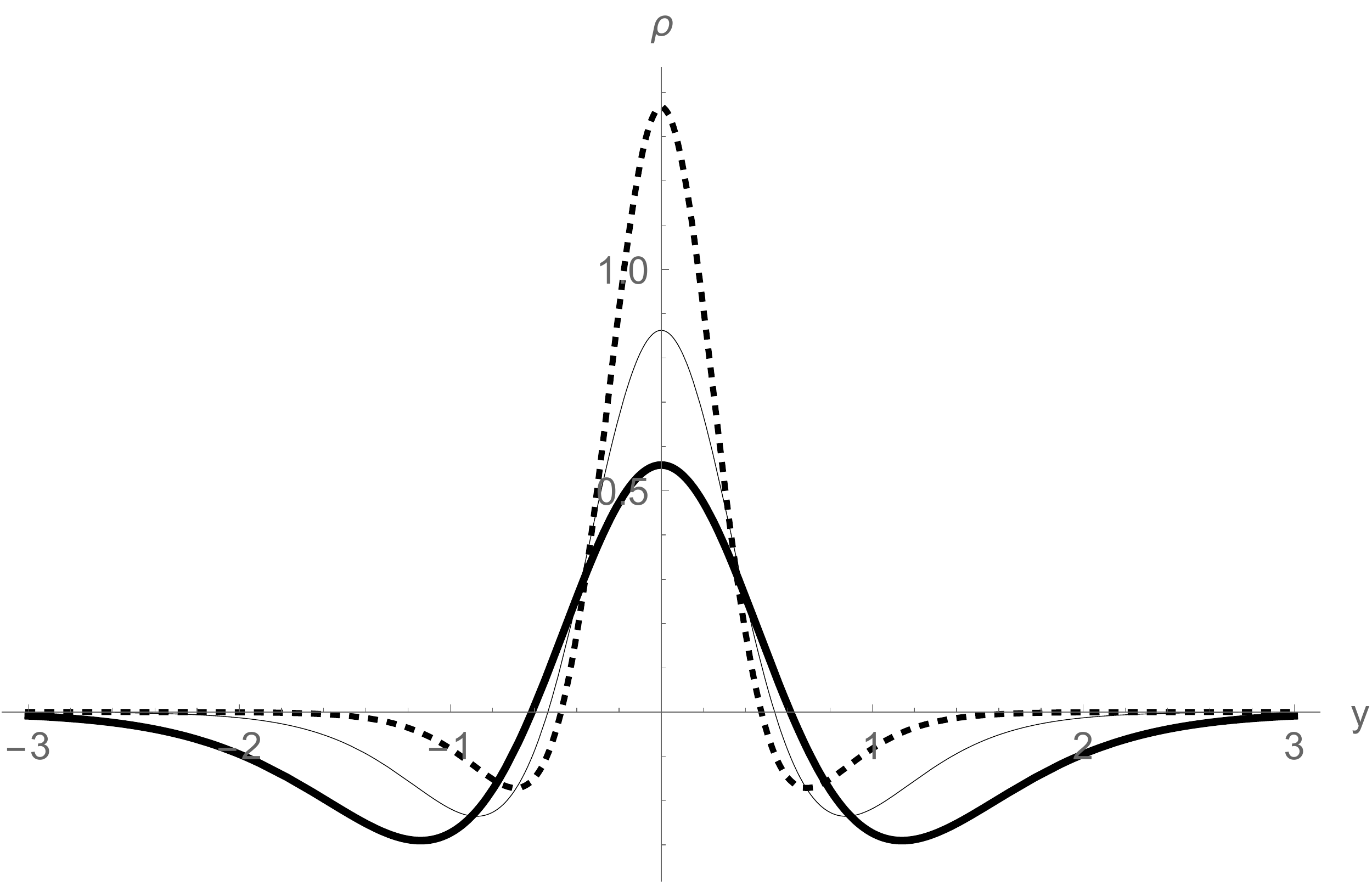}
    \caption{Energy density for $b=c=1$. As the parameter $\alpha$ increases the brane width decreases.}
    \label{energy density}
\end{figure}

As shown in Fig.\ref{energy density}, the Q-gravity controls the width and the amplitude of the energy density. However, the overall properties are preserved compared to the GR sine-{Gordon} model \cite{Gremm}.

\subsection{Non-quadratic gravity $k\neq 0$}

We will now investigate the effects of additional non-quadratic gravitational dynamics on the braneworld. For this purpose, we will first analyse the modifications on the vacuum solution (exterior). The gravitational equations lead to
\begin{equation}
    (\sigma_1 -2\sigma_0)(f_{\mathbb{Q}}A'' + f'_{\mathbb{Q}}A')-8\sigma_0 f_{\mathbb{Q}} A'^{2}=0.
\end{equation}
Assuming this conditions and considering $\Lambda\neq 0$ and $A'=- c$, the gravitational equations with cosmological constant lead to
\begin{equation}
\label{cLambdaequation}
    8\sigma_1 c^2 + k\left((n-1/2)\Big[16\sigma_1c^{2}\Big]^{n}\right)=-\Lambda 
\end{equation}
Curiously, if $n=\frac{1}{2}$, then Eq.(\ref{cLambdaequation}) coincides with Eq.(\ref{qgravitythinlimit}), regardless of the value of $k$. Thus, the power $n=1/2$ provides no correction to the geometry outside the 3-brane for $\Lambda\neq 0$. For $n=2$,  Eq.(\ref{cLambdaequation}) admits the solution 
\begin{eqnarray}
\label{n=2rs}
\sigma_1c^2=\frac{ \sqrt{1-192k\Lambda}-1}{96k} ,
\end{eqnarray}
which for small values of the product $k\Lambda$ can be approximated as $\sigma_1c^2\approx \Lambda$. 
Eq.(\ref{n=2rs}) shows how a thin RS braneworld \cite{rs1} is modified by the non-quadratic $f(\mathbb{Q})$ gravity. {Likewise the quadratic solution in Eq.(\ref{qgravitythinlimit}), both $AdS_5$ and $dS_5$ bulk spacetimes are allowed by Eq.(\ref{n=2rs}).} For a $dS_5$ bulk, Eq.(\ref{n=2rs}) constrains the value of the parameter $k$ to the interval $k\leq 1/192\Lambda$.  

Now let us consider thick 3-brane solution for $\Lambda = 0$.  We seek for BPS solutions in the non-quadratic dynamics, by considering a superpotential $W(\phi)$ such that \cite{Gremm,wilami,Menezes}
\begin{equation}
\label{bps1}
    A'=-\frac{1}{3}W(\phi)
\end{equation}
For 
$\kappa_5 = 2$, the gravitational equations and the BPS equation \eqref{bps1} yield 
\begin{equation}
\label{bps2}
    \phi' = \frac{W_\phi}{2}\left(1+k(C_n +D_n ) W^{2n-2}\right),
\end{equation}
where $C_n = n 2^{2(n-1)}3^{1-n}$ and $D_n =n(n-1)(\sigma_1 - 2\sigma_0)2^{2n+1}3^{1-n}$. Note that, for $n=1/2$, the usual GR-based BPS equations are recovered \cite{Gremm, wilami, Menezes}. Using Eq.\eqref{bps1} and Eq.\eqref{bps2} the Eq.\eqref{eom1} leads to
\begin{eqnarray}
\label{bps3}
    V(\phi)&=&\frac{W_{\phi}^{2}}{8}\left(1+k C_n W^{2n-2}\right) \Bigg[1+k(C_n +D_n) W^{2n-2}\Bigg]\nonumber\\
    &-& \frac{W^2}{3}\Bigg[1+kC_n \left(2-\frac{1}{n}\right) W^{2n-2}\Bigg]
\end{eqnarray}
The BPS equations \eqref{bps1},\eqref{bps2} and \eqref{bps3} reduce the second order EOM into a system of first-order equations. Note that the modified gravitational dynamics also influences the scalar field profile and the potential properties. Similar BPS equations were found in teleparallel $f(T)$ \cite{Yang2012,Menezes,ftnoncanonicalscalar}, $f(T,B)$ \cite{allan1} and $f(Q)$ \cite{fqbrane}.

Let us see how the $f(\mathbb{Q})$ dynamics modifies the scalar field features in the well-known sine-gordon model, where $    W(\phi)=3bc \sin{\Bigg[\sqrt{\frac{2}{3b}}\phi\Bigg]}.$

\begin{figure}
\begin{center}
\begin{tabular}{ccc}
\includegraphics[height=5cm]{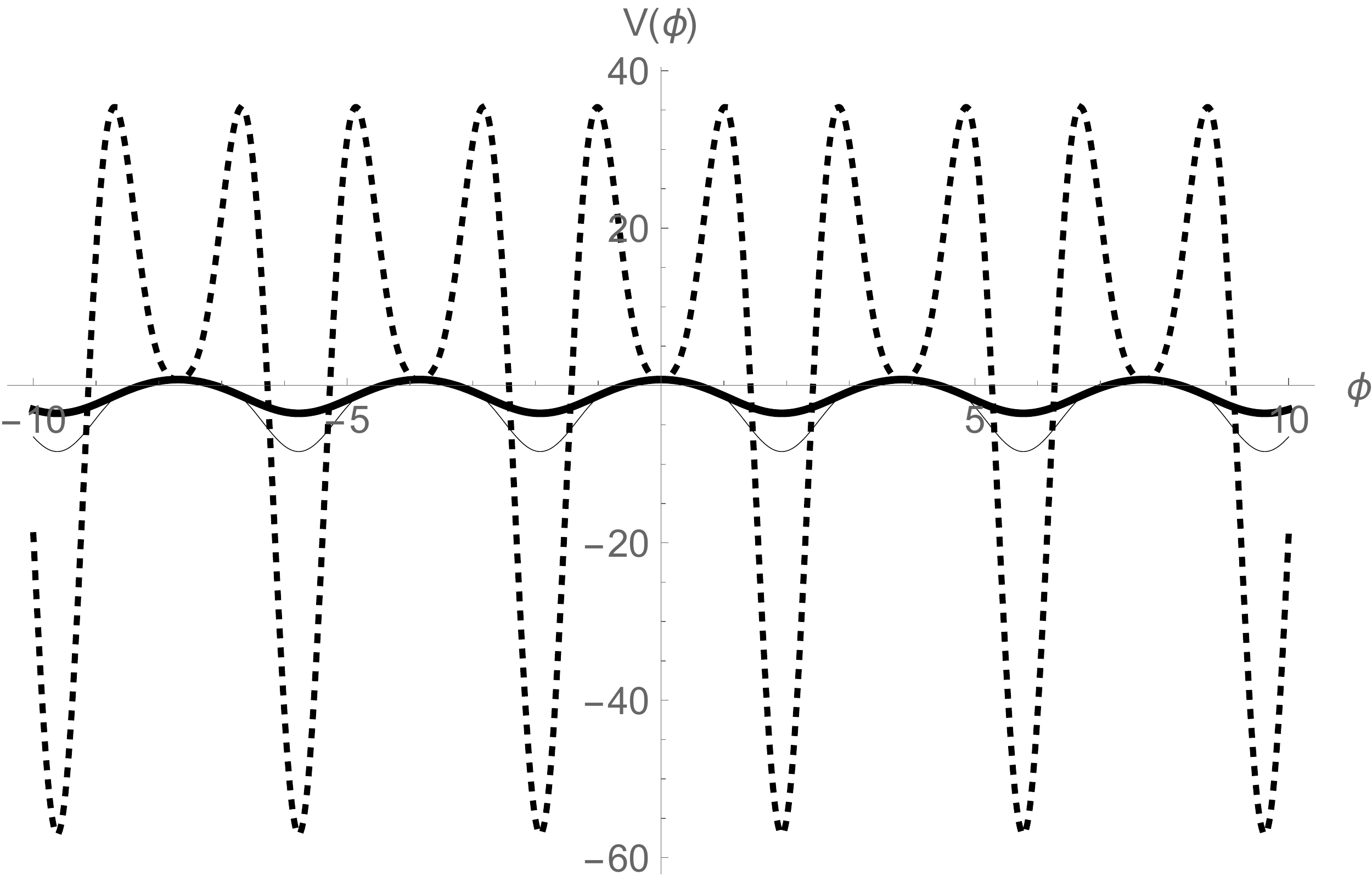}
\includegraphics[height=5cm]{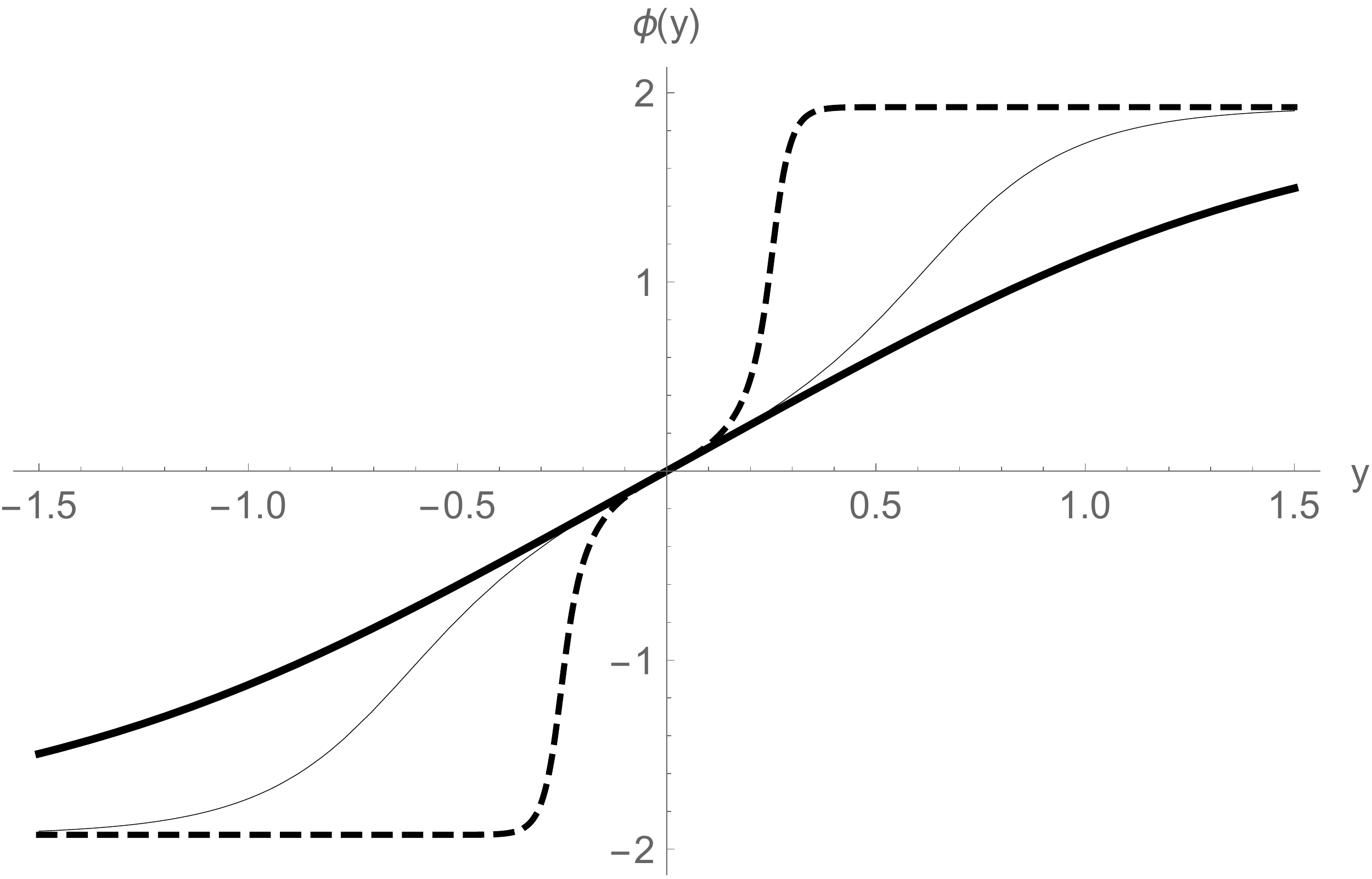}\\ 
(a) \hspace{8 cm}(b)
\end{tabular}
\end{center}
\caption{Potential (a) and scalar field (b) for $n=2$ and $k=0.005$ (thick line), $k=0.05$ (thin line), $k=0.5$ (dashed line).}
\label{fig2}
\end{figure}

In fig.\ref{fig2} (a) we plotted the potential for $n=2$ at the GR point $\sigma_0=$ and $\sigma_1=3/4$. As $k$ increases the potential is modified. Indeed, the wells become deeper and the initial barriers become false vaccua. In fig.\ref{fig2} (b) we plotted the respective scalar field solution. The increasing of $k$ leads to the tendency of forming a plateau around the origin. Moreover, the field tends to attain the vacuum in a finite distance. These features reveal that, despite the presence of only one scalar field, the modified Q-gravity dynamics produces a thick brane with an inner structure resembling a hybrid defect \cite{compacton}. Similar results were found in $f(T)$ \cite{Menezes} and $f(T,B)$ \cite{allan1} braneworlds.

\begin{figure}
\begin{center}
\begin{tabular}{ccc}
\includegraphics[height=5cm]{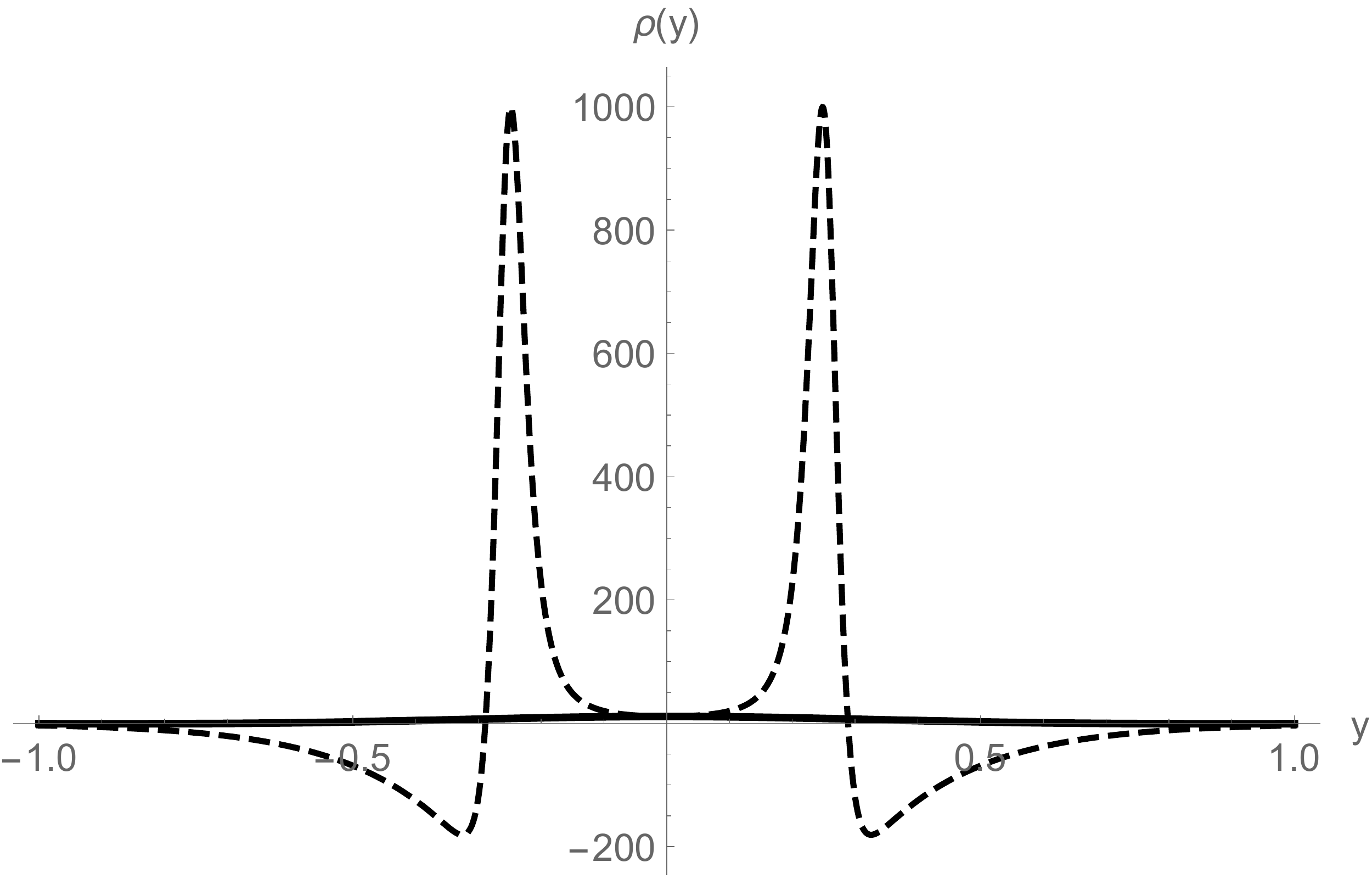}
\includegraphics[height=5cm]{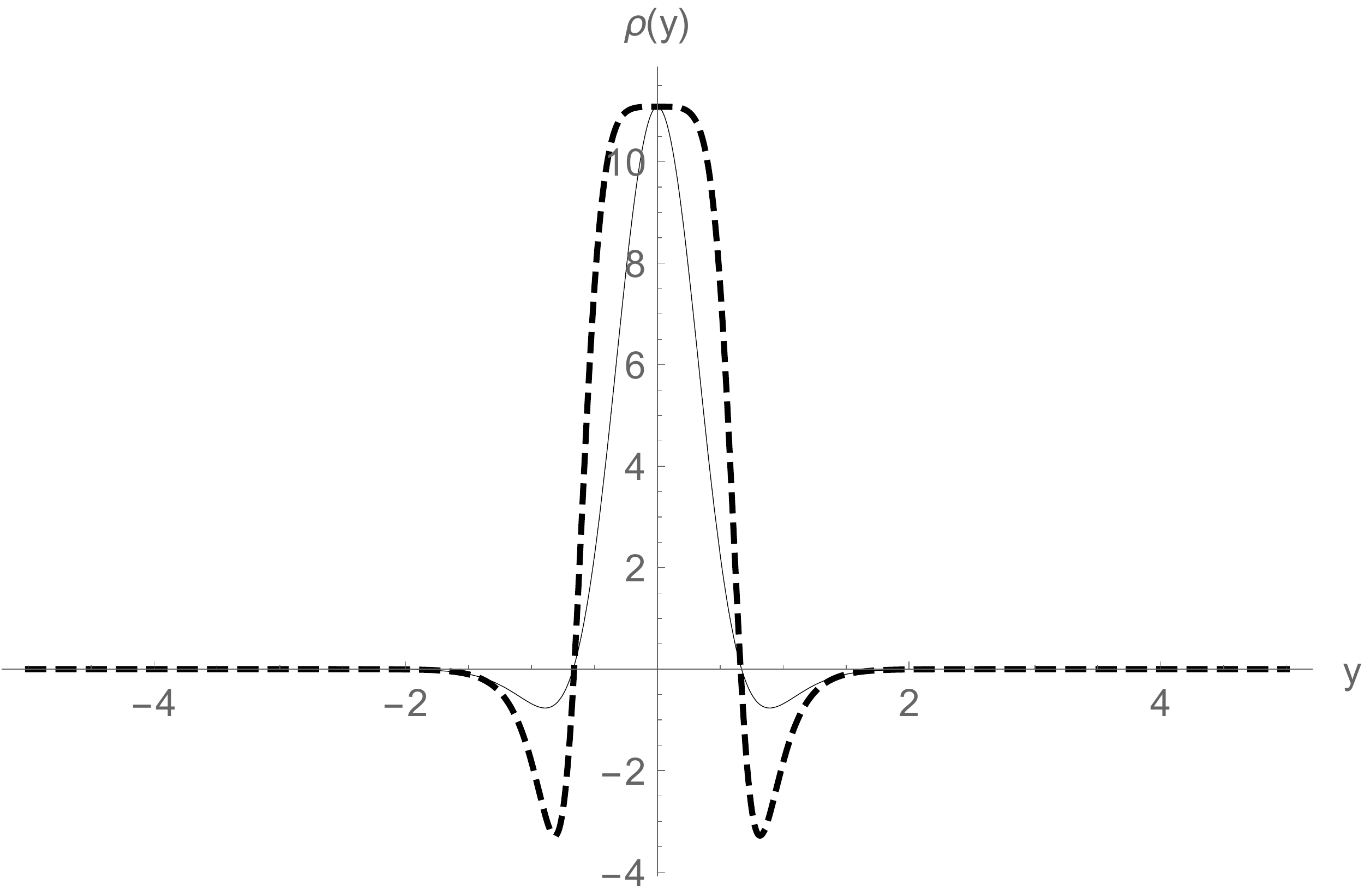}\\ 
(a) \hspace{8 cm}(b)
\end{tabular}
\end{center}
\caption{Plots of the energy density for $n=2$. In (a) for $k=0.5$ (dashed line). In (b) for $k=0.005$ (thin line) and for $k=0.05$ (dashed line).}
\label{fig3}
\end{figure}

In fig.\ref{fig3} (a) we plotted the energy density. This figure reveals another interesting effect. In (a) the energy density exhibits two distinctive peaks revealing the brane splitting. Figure \ref{fig3} (b) shows that the transition between the single to two brane is performed maintaining the geometry smooth. Moreover, the increasing of $k$ leads to a plateau around the origin. Therefore, as $k$ increases the brane undergoes a smooth transition from one to two branes.

\begin{figure}
\begin{center}
\begin{tabular}{ccc}
\includegraphics[height=5cm]{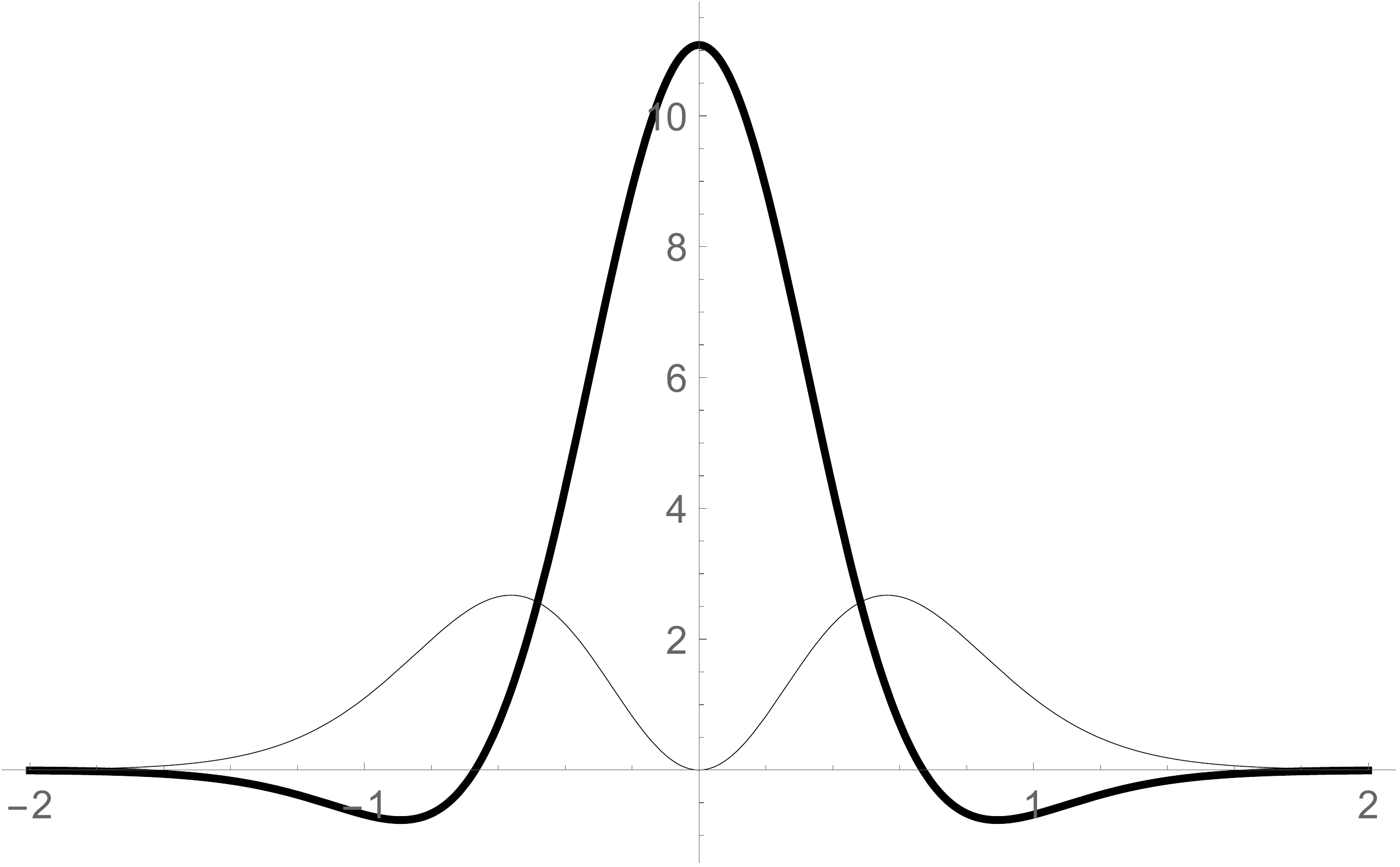}
\includegraphics[height=5cm]{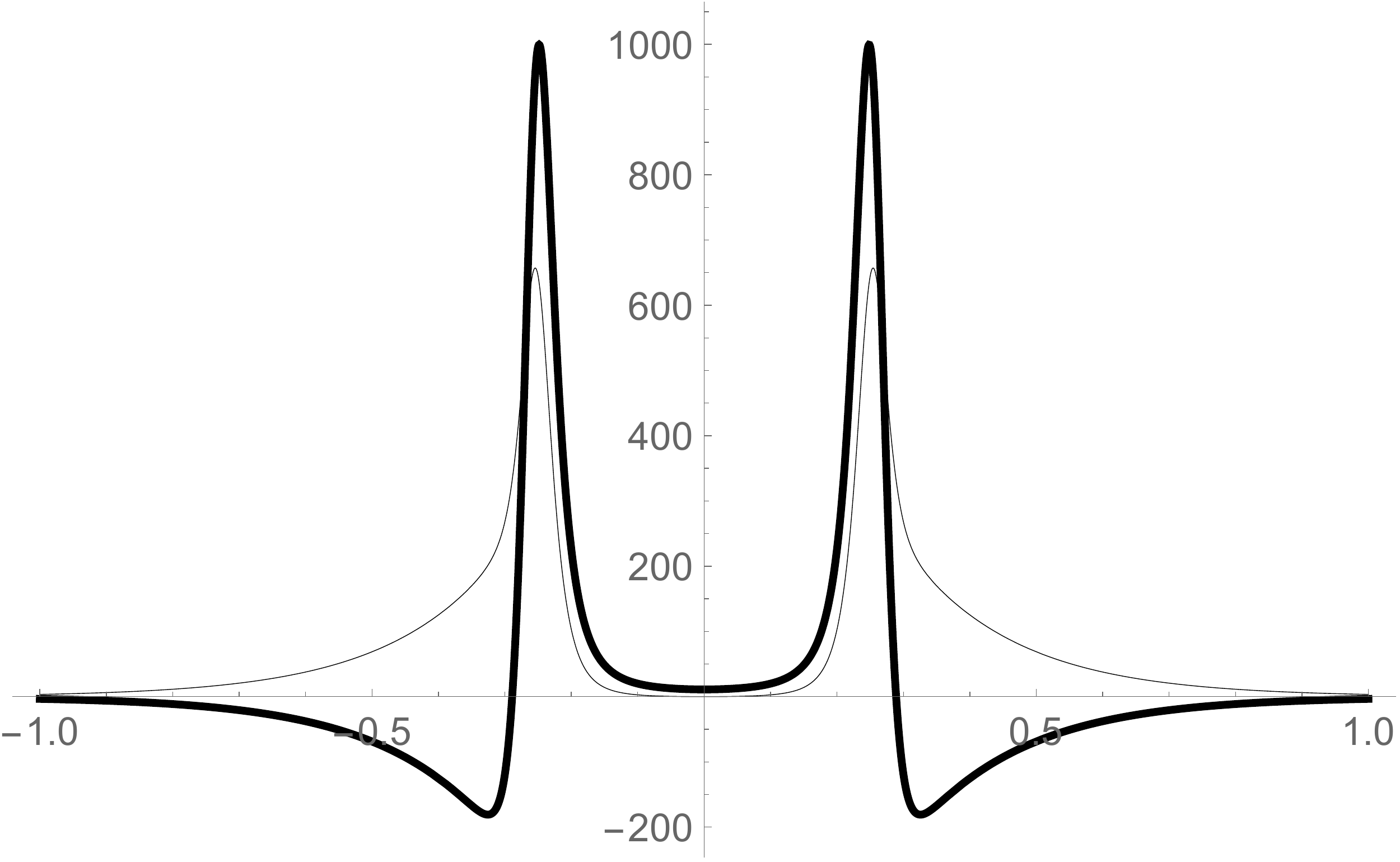}\\ 
(a) \hspace{8 cm}(b)
\end{tabular}
\end{center}
\caption{Plots of the energy density (thick line) and the pressure (thin line) for (a) $k=0.005$  and (b) for $k=0.5$.}
\label{fig4}
\end{figure}

In figure \ref{fig4} we plot the energy density (thick line) and the pressure (thin line) for $k=0.005$ in (a) and for $k=0.5$ in panel (b). For small $k$ the energy density exhibits a symmetric peak around the origin whereas the pressure has two peaks displayed from the origin. As $k$ increases, the energy density also develops two peaks, a hallmark of the brane splitting. It is worthwhile to mention that, im most of the point, the source satisfies the weak energy condition $\rho \geq 0$. In addition, the strong energy condition $\rho + P \geq 0$ is satisfied and around the two peaks (brane core) the dominat energy condition $\rho \geq P$. 

\section{Fermions}
\label{sec4}
After studying the gravitational field and the bulk geometry for the quadratic $\mathbb{Q}$ and modified $\mathbb{Q}+k\mathbb{Q}^2$, we now investigate the features of fermionic matter in these modified gravitational scenarios.  We propose a non-minimal coupling of a massless bulk spinor and the gravitational field in the form: 
\begin{equation} \label{fermion-action}
S_\Psi = \int {d^{5} x \sqrt{-g} \Big[\bar{\Psi} i \Gamma ^{M} D_{M} \Psi - \beta \sqrt{\mathbb{Q}}\bar{\Psi}\Psi\Big]},
\end{equation}
where $\beta$ is a dimensionless non-minimal coupling constant, $\Gamma^{M} = e_{\bar{M}}^{M} \gamma^{\bar{M}}$ are the Dirac matrices in a curved spacetime, $\gamma^{\bar{M}}$ are the flat Dirac matrices and the  \textit{vielbeins} $e_{\bar{M}}^{M}$ satisfy 
$g_{M N} = \eta _{\bar{M}\bar{N}}e_{M}^{\bar{M}} e_{N}^{\bar{N}}$. The spinor covariant derivative in nonmetricity spacetimes is given by \cite{delhom}
\begin{equation} \label{deri-covar}
D_{M} = \partial _{M} + \Omega_{M}^{LC},
\end{equation}
where $\Omega_{M}^{LC}  =\frac{1}{4} \Gamma_{M} ^{\bar{M} \bar{N}} \gamma _{\bar{M}} \gamma _{\bar{N}}$ is the torsion-free spinor connection and $\Gamma_{M} ^{\bar{M} \bar{N}}$ are 1-form Levi-Civita connection coefficients, i.e., 
$\omega_{\bar{N}}^{\bar{M}}=\Gamma_{M\bar{N}}^{\bar{M}}dx^{M}$. As pointed out in Ref.\cite{delhom}, in non-Riemaniann spacetimes the non-vanishing term added to the Riemaniann covariant derivative comes from the torsion alone. Thus, in the symmetric teleparallel gravity, the spinor connection structure is preserved.

In order to confine the spin-1/2 massless mode on the 3-brane, a Yukawa-like interaction with an additional scalar field is usually adopted. In this work we consider a non-minimal coupling in Eq.(\ref{fermion-action}) which leads to the localization of the massless mode by gravitational interactions only. Similar non-minimal couplings involving torsion scalars were studied in
a teleparallel \cite{allan} and in the Lyra geometry \cite{lyra} braneworlds.

In the conformal coordinate $z=\int{e^{-A}dy}$, the metric can be written as $ds^2=\eta_{\bar{M}\bar{N}}\hat{\theta}^{\bar{M}}\otimes \hat{\theta}^{\bar{N}}$, where $\hat{\theta}^{\bar{M}}=e^{-A}\delta^{\bar{M}}_{M}dx^{M}$. The torsion-free condition $d\hat{\theta}^{\bar{M}}+\omega^{\bar{M}}_{\bar{N}}\hat{\theta}^{\bar{N}}=0$ leads to the nonvanishing 1-form connection coefficients $\Gamma^{\bar{\mu}}_{5\bar{\nu}}=\dot{A}\delta^{\bar{\mu}}_{\bar{\nu}}$ and $\Gamma^{5}_{\bar{\mu}\bar{\nu}}=\dot{A}\eta_{\bar{\mu}\bar{\nu}}$, where the dot stands for the derivative with respect to the conformal coordinate, i.e., $d/dz$.  Accordingly, the Dirac equation takes the form
\begin{equation}
\Big[\gamma^\mu \partial_\mu +\gamma^z(\partial_z + 2\dot{A})-e^{A}(\beta  \sqrt{\mathbb{Q}})\Big]\Psi=0.
\end{equation}
Let us perform the Kaluza-Klein decompostition $\Psi(x,z)=f(z)\sum{\psi_{4R} (x)\psi_R (z) +\psi_{4L}\psi_L (z)}$, where $f(z)=e^{2A}$ and $\psi_{R,L}$ are, respectively, the right-handed and left-handed chiral states with respect to the extra dimension. Assuming that the spin-1/2 fermion satisfies the on-brane Dirac equation $\gamma^\mu \partial_\mu \Psi=m\Psi$, we obtain 
\begin{eqnarray}
\label{coupleddiracequations}
(\partial_z +  e^{A}\beta \sqrt{\mathbb{Q}})\psi_L &=& m \psi_R\nonumber\\
(\partial_z - e^{A}\beta\sqrt{\mathbb{Q}})\psi_R &=&- m \psi_L.
\end{eqnarray} 
It is worthwhile to mention that the $\dot{A}$ term stemming from the connection was absorbed by a field redefinition denoted by the function $f(z)$ \cite{fermionsinegordon,casa}. 

Decoupling the Dirac system in Eq.(\ref{coupleddiracequations}), we obtain Schr\"{o}dinger-like equations for each chirality, as
\begin{equation}
\label{schrodingerequation}
 -\ddot{\psi}_{R,L}+U_{R,L}(z)\psi_{R,L}=m^2\psi_{R,L},   
\end{equation}
where the Schr\"{o}dinger-like potential $U_{R,L}$ is given by
\begin{equation}
\label{spinorpotential}
U_{R,L}= W^2 \pm \dot{W},
\end{equation}
and $W(z)=\beta e^{A}\mathbb{Q}$ is the so-called superpotential of the SUSY-like quantum mechanics potential in Eq.(\ref{spinorpotential}). The Schr\"{o}dinger-like Eq.(\ref{schrodingerequation}) determines the KK states and their corresponding massive spectrum. The SUSY-like structure of the potential in Eq.(\ref{spinorpotential}) guarantees that $m^2 \geq 0$, thus avoiding tachionic KK states \cite{davi}. In addition, the SUSY-like structure also allows for the existence of a massless mode $m=0$
in the form \cite{fermionsinegordon,casa,davi}
\begin{equation}
    \psi_0 = e^{-\int{Wdz}}.
\end{equation}
Next we will study the properties of the spin-1/2 fermion in two different regimes.

\subsection{Quadratic gravity limit}

\begin{figure}[htb] 
       \begin{minipage}[b]{0.48 \linewidth}
           \includegraphics[width=\linewidth]{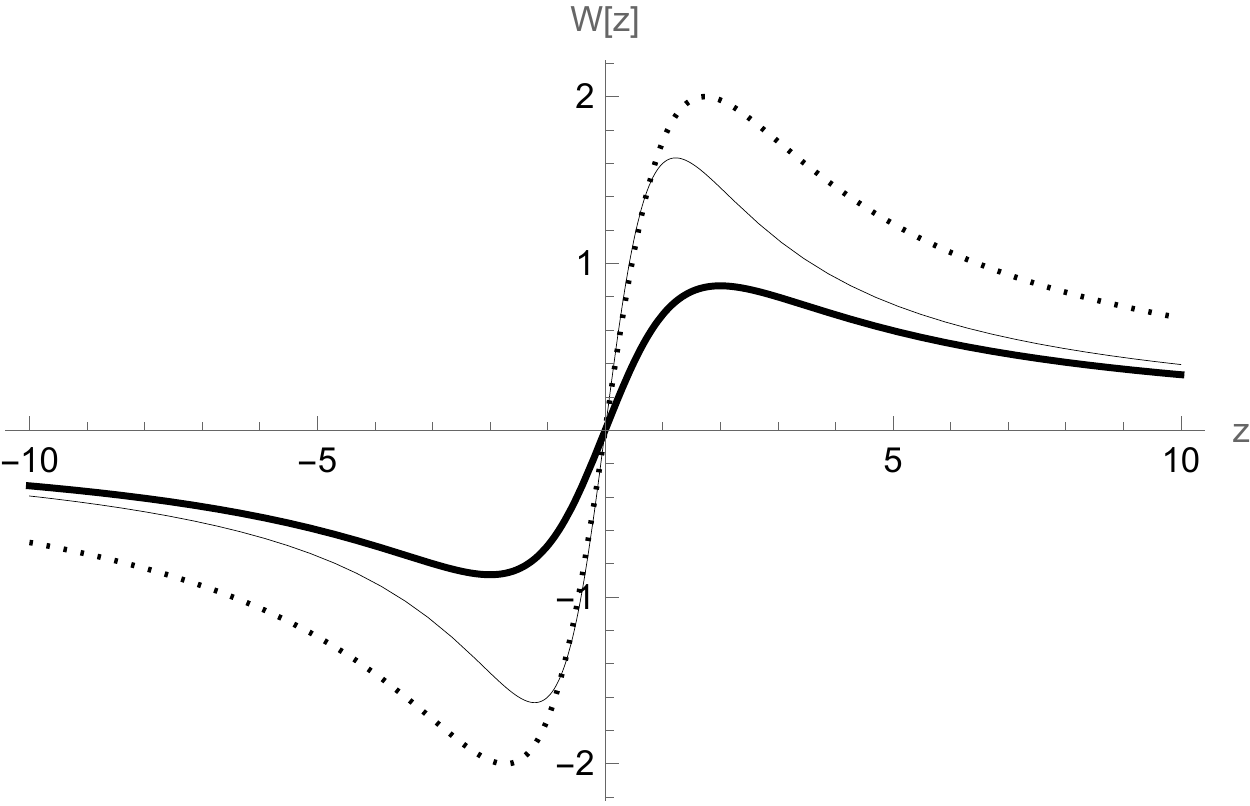}\\
           \caption{Fermion superpotential.}
          \label{superpotential}
       \end{minipage}\hfill
       \begin{minipage}[b]{0.48 \linewidth}
           \includegraphics[width=\linewidth]{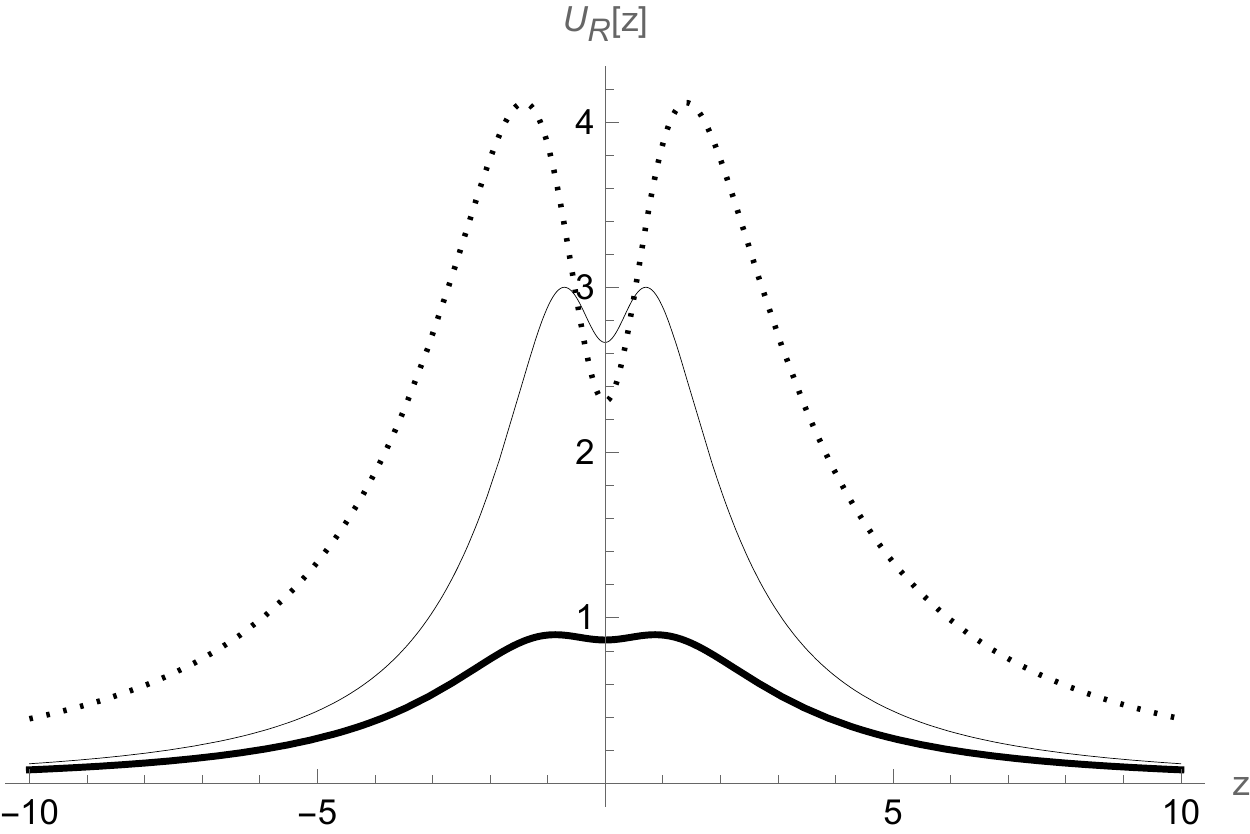}\\
           \caption{Fermion potential $U_R$.}
           \label{rightpotentialdisplay}
       \end{minipage}
\end{figure}

\begin{figure}[htb] 
       \begin{minipage}[b]{0.48 \linewidth}
           \includegraphics[width=\linewidth]{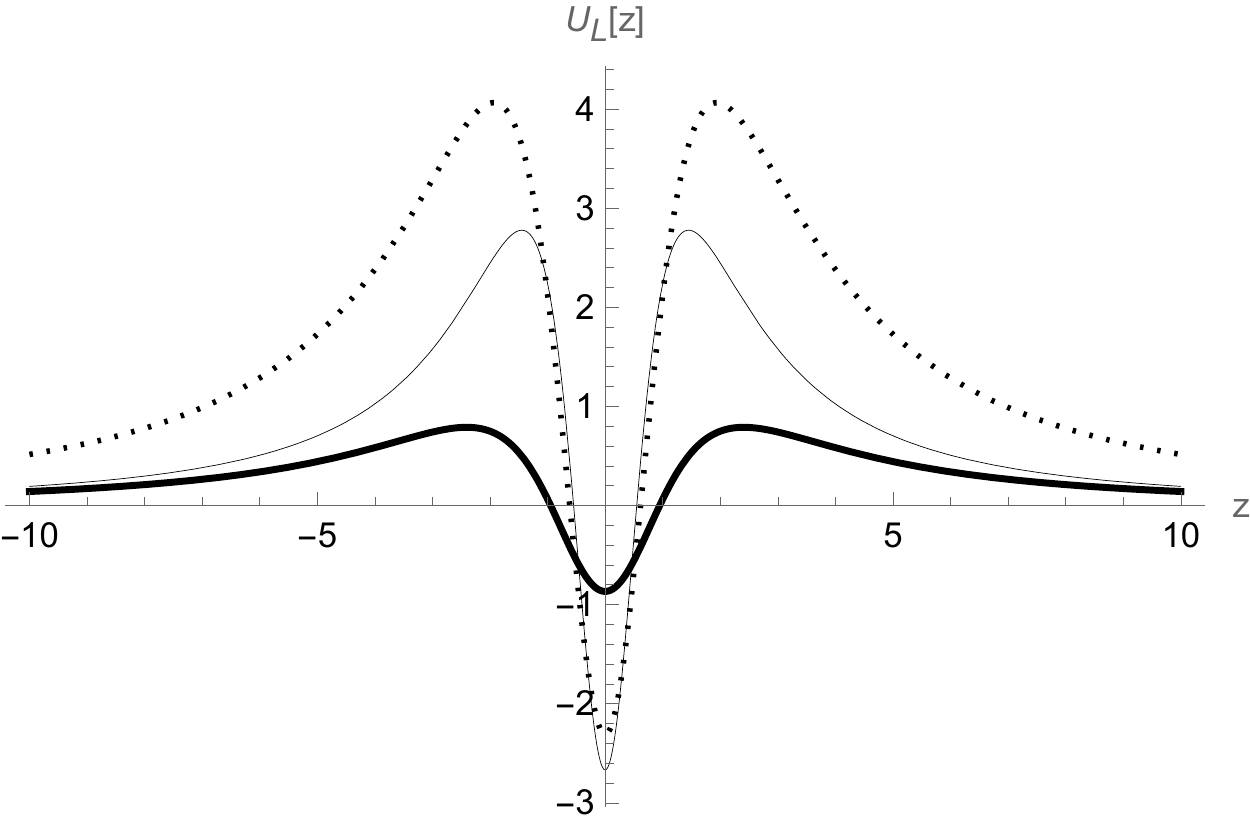}\\
           \caption{Fermion potential $U_L$.} 
          \label{quadraticleftpotentialfigure}
       \end{minipage}\hfill
       \begin{minipage}[b]{0.48 \linewidth}
           \includegraphics[width=\linewidth]{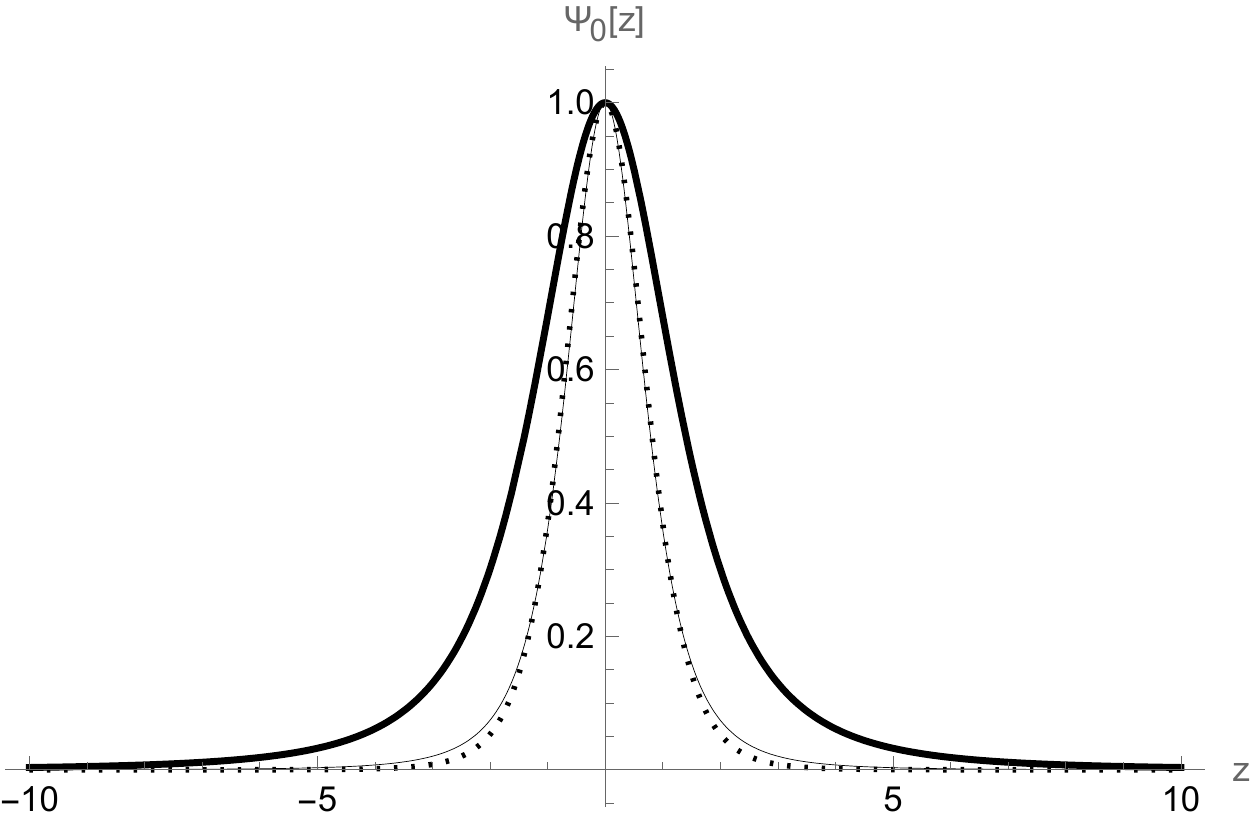}\\
           \caption{Left-handed massless mode $\Psi_0$.}
           \label{quadraticzeromodefigure}
       \end{minipage}
\end{figure}

Considering $k=0$ in the gravitational Lagrangian, the warp function is given by Eq.\ref{quadraticwarpedfunction}. The corresponding conformal coordinate,  $z=\int{\cosh(\lambda y)^{b/\sqrt{8\alpha}}dy}$, can only be analytically inverted when 
$b/\sqrt{8\alpha}=1$, wherein $z=\frac{1}{\lambda}\sinh(\lambda y)$. 
In this case, the warped function has the expression
\begin{equation}
    A(z)=\ln(1+\lambda^2 z^2)^{-1/2},
\end{equation}
where $\lambda=\sqrt{2\alpha}c$. As a result, the superpotential has the form
\begin{equation}
    W(z)=-\frac{4\beta\lambda^2\sqrt{ \sigma_1}z}{1+\lambda^2 z^2},
\end{equation}
which exhibits the behaviour shown in the fig.(\ref{superpotential}). The left-handed potential is given by
\begin{equation}
    U_L = -\frac{4\lambda^2\sqrt{ \sigma_1}(1-4\sigma_1\lambda^2 z^2 )}{(1+\lambda^2 z^2)^2}.
\end{equation}
Likewise the gravitational field, the potential above depends explicitly only on the effective parameters $\sigma_1$ and $\sigma_0$ (via $\alpha$).
The left-handed potential is plotted in fig.(\ref{quadraticleftpotentialfigure}), where it shows the usual volcano shape, similar to those models with Yukawa interaction \cite{fermionsinegordon,casa}. Using the condition $b^2=8\alpha$ and since $\lambda=\sqrt{2\alpha}c$, the values chosen for the $c_i$ coefficients have to satisfy $\sigma_1 + 2\sigma_0 >0$. We seek for modified solutions around the GR-based geometry. Thus, we plotted the potential for the GR equivalent case, $\sigma_0 =0$ and $\sigma_1 = 3/4$, corresponding to $c_1=-c_3 =-\frac{1}{4}$, $c_5=-1/2$ (thick line), $\sigma_0 =1/2$, $\sigma_1 = 1$, corresponding to $c_1=0$, $c_3=1/4$,  $c_5=\frac{2\sqrt{3}}{3}$ (thin line), and $\sigma_0 =-1$, $\sigma_1 = 3$, corresponding to $c_1=2$, $c_3 = 1/4$ and $c_5=-\frac{3}{2}$ (dashed line). The graphic for the superpotential (Fig.\ref{superpotential}), right-handed potential (Fig.\ref{rightpotentialdisplay}) and the left-handed massless mode (\ref{quadraticzeromodefigure}) were plotted for the same values discussed above.  The minimum of the potential is controlled mainly by the $\sigma_1$ coefficient according to $U_{R0}=4\lambda^2 \sqrt{ \sigma_1}$, where some dependence on $\sigma_0$ also exists. Therefore, by varying $\sigma_1$, it is possible to vary the heigh of the barriers and the depth of the potential well. Moreover, the asymptotic behaviour of the potential indicates a continuous tower of massive non-localized KK modes. 

The corresponding massless mode is given by
\begin{equation}
    \psi_0 = N(1+\lambda^2 z^2)^{-2\sqrt{ \sigma_1}},
\end{equation}
whose behaviour is sketched in Fig.(\ref{quadraticzeromodefigure}). The massless mode is a bound state to the 3-brane and its behaviour is only slightly modified by the variation of the $c$ coefficients. 

\subsection{Non-quadratic regime $n=2$}

\begin{figure}[htb] 
       \begin{minipage}[b]{0.48 \linewidth}
           \includegraphics[width=\linewidth]{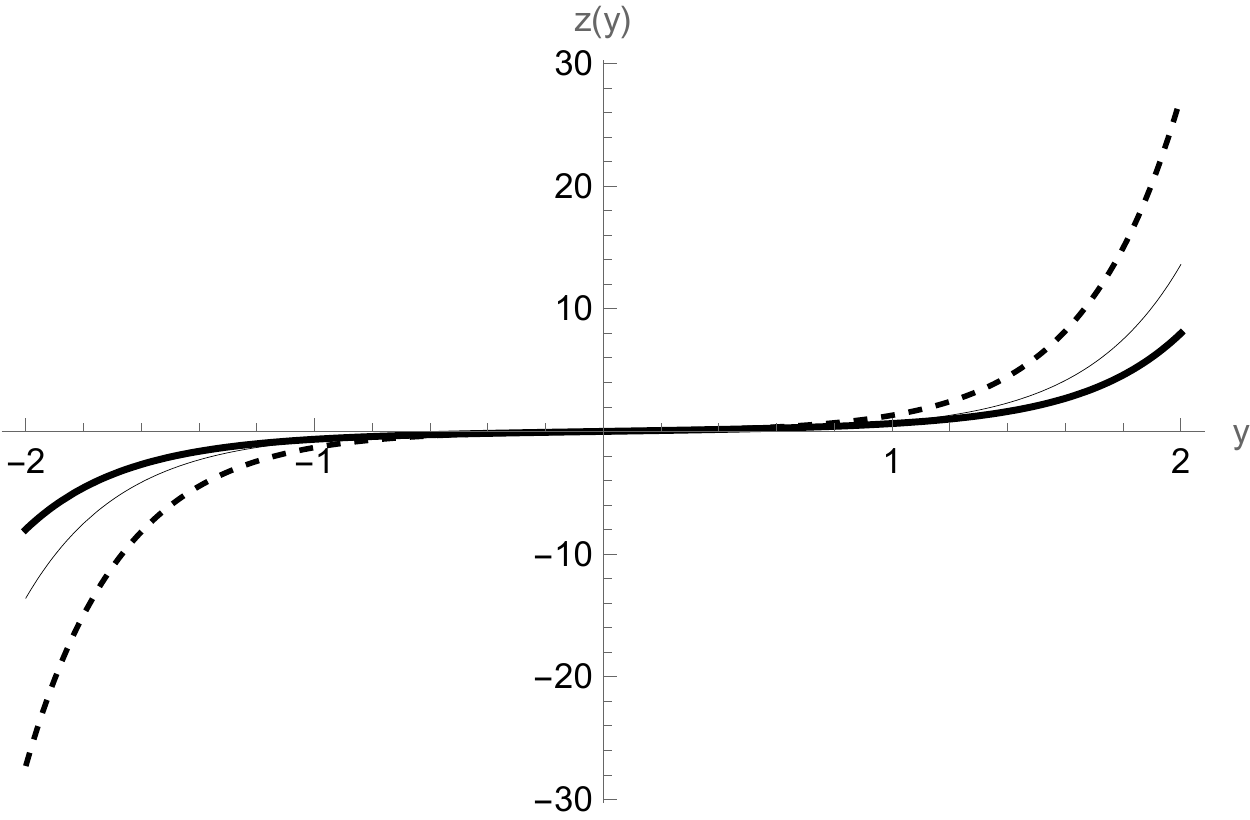}\\
           \caption{Conformal coordinate $z(y)$ for $b=c=1$. For $k=0.005$ (thick line), $k=0.05$ (thin line) and $k=0.5$ (dashed line), $z(y)$ is a smooth one to one function.} 
          \label{conformalcoordinate}
       \end{minipage}\hfill
       \begin{minipage}[b]{0.48 \linewidth}
           \includegraphics[width=\linewidth]{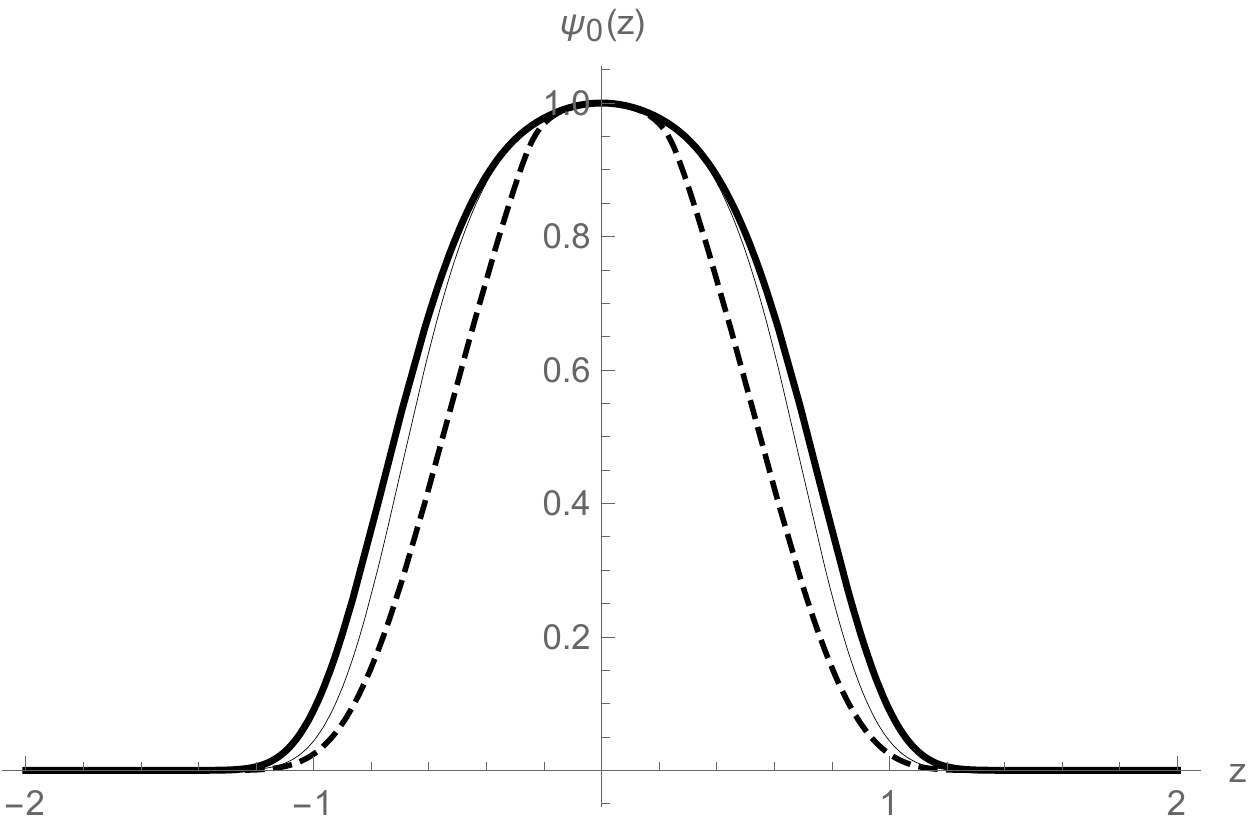}\\
           \caption{Left-handed massless mode $\alpha_0$ for $b=c=\beta=1$. Despite the split undergone by the brane, this KK ground state keeps the usual bell-shape.}
           \label{nonquadraticmasslessmode}
       \end{minipage}
\end{figure}

When the parameter that controls the geometric corrections $k Q^n$ is nonzero, the equations get very messy and one must resort to numerical methods in order to solve for the warp factor $A(y)$ and the scalar field $\phi(y)$. The Schr\"{o}dinger-like potentials $U_{R,L}$ and the massless mode $\alpha_0$ can only be obtained and analysed numerically, as well. To illustrate the changes, Fig.(\ref{conformalcoordinate}) shows the behaviour of $z(y)$, where a smooth change from $y$ to $z$ is shown. Fig.(\ref{nonquadraticmasslessmode}) depicts the behavior of the left-handed massless mode $\psi_0$. Remarkably, even though the thick 3-brane undergoes a split transition driven by the change in the parameter $k$, the massless mode remains peaked around the origin. All the figures correspond to the GR-like configuration $\sigma_0 = 0$ and $\sigma_1 =3/4$.   

\begin{figure}[htb] 
       \begin{minipage}[b]{0.48 \linewidth}
           \includegraphics[width=\linewidth]{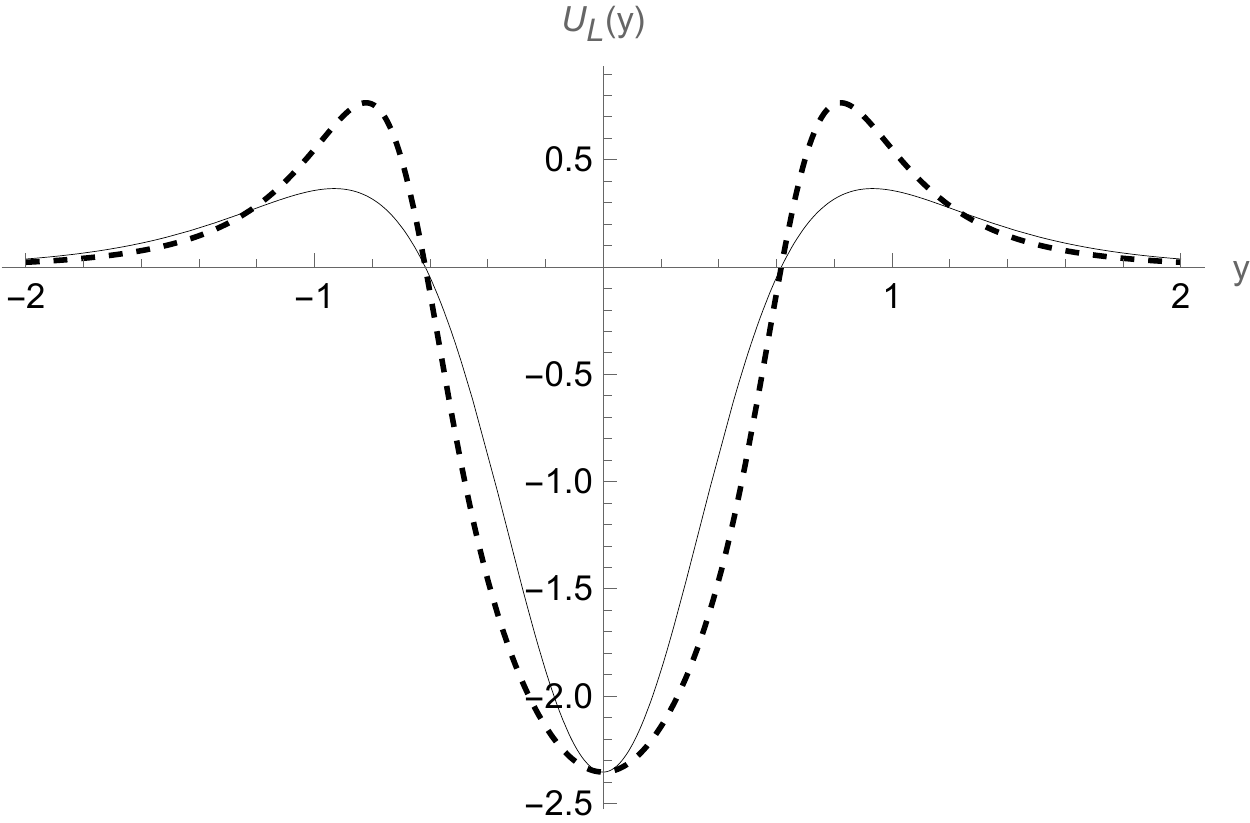}\\
           \caption{Left-handed potential $U_L$ for $b=c=\beta=1$. For $k=0.005$ (thin line) the potential has the volcano-like shape. For $k=0.05$ (dashed line) the barriers increase their heigh and the potential well becomes wider.} 
          \label{nonquadraticleftpotentialfigure}
       \end{minipage}\hfill
       \begin{minipage}[b]{0.48 \linewidth}
           \includegraphics[width=\linewidth]{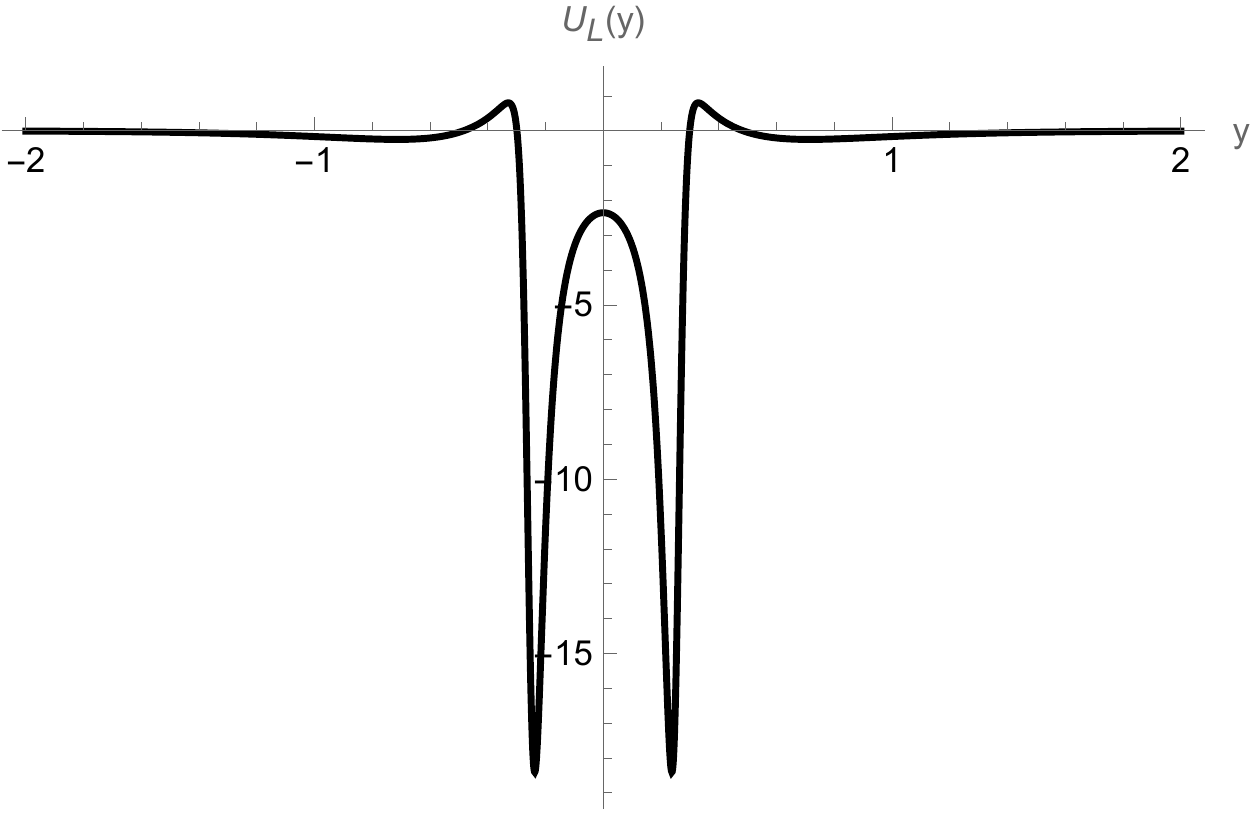}\\
           \caption{Left-handed potential for $\lambda=1$ and $k=0.5$. The potential still vanishes asymptotically. However, the potential well undergoes a transition leading to a barrier at the origin, followed by two symmetric and deep potential wells at both sides of the origin.}
           \label{nonquadraticpotential}
       \end{minipage}
\end{figure}

The Schr\"{o}dinger-like potential $U_L$, on the other hand, undergoes a strong transition. As shown in fig.(\ref{nonquadraticleftpotentialfigure}), for $k=0.005$ (thin line) and $k=0.05$ (dashed line), only the heigh of the barrier and the width of the potential well do slightly vary. However, for $k=0.5$, as shown in fig.(\ref{nonquadraticpotential}), the potential well turns into a barrier at the origin, followed by two deep, symmetrical and thin wells around the origin.

\begin{figure}[htb] 
       \begin{minipage}[b]{0.48 \linewidth}
           \includegraphics[width=\linewidth]{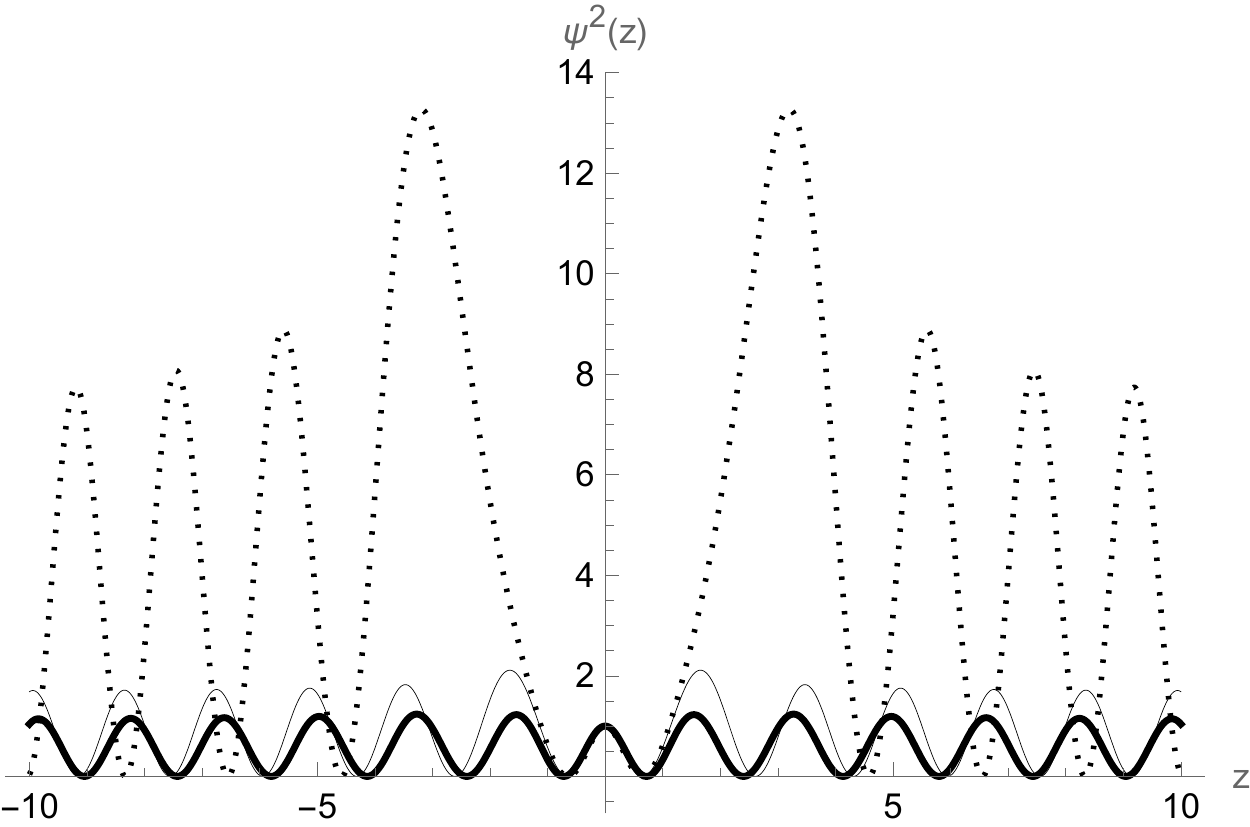}\\
           \caption{Left-handed massive mode for $m=4$. For $\sigma_0 =0, \sigma_1=3/4$ (thick line) and for $\sigma_0 =1/2, \sigma_1=1$ (thin line), the amplitude of this massive mode varies only a small fraction around the origin. For $\sigma_0 =-1, \sigma_1=3$, there are two symmetric peaks around the origin and the massive mode rapidly decays.} 
          \label{quadraticmassive}
       \end{minipage}\hfill
       \begin{minipage}[b]{0.48 \linewidth}
           \includegraphics[width=\linewidth]{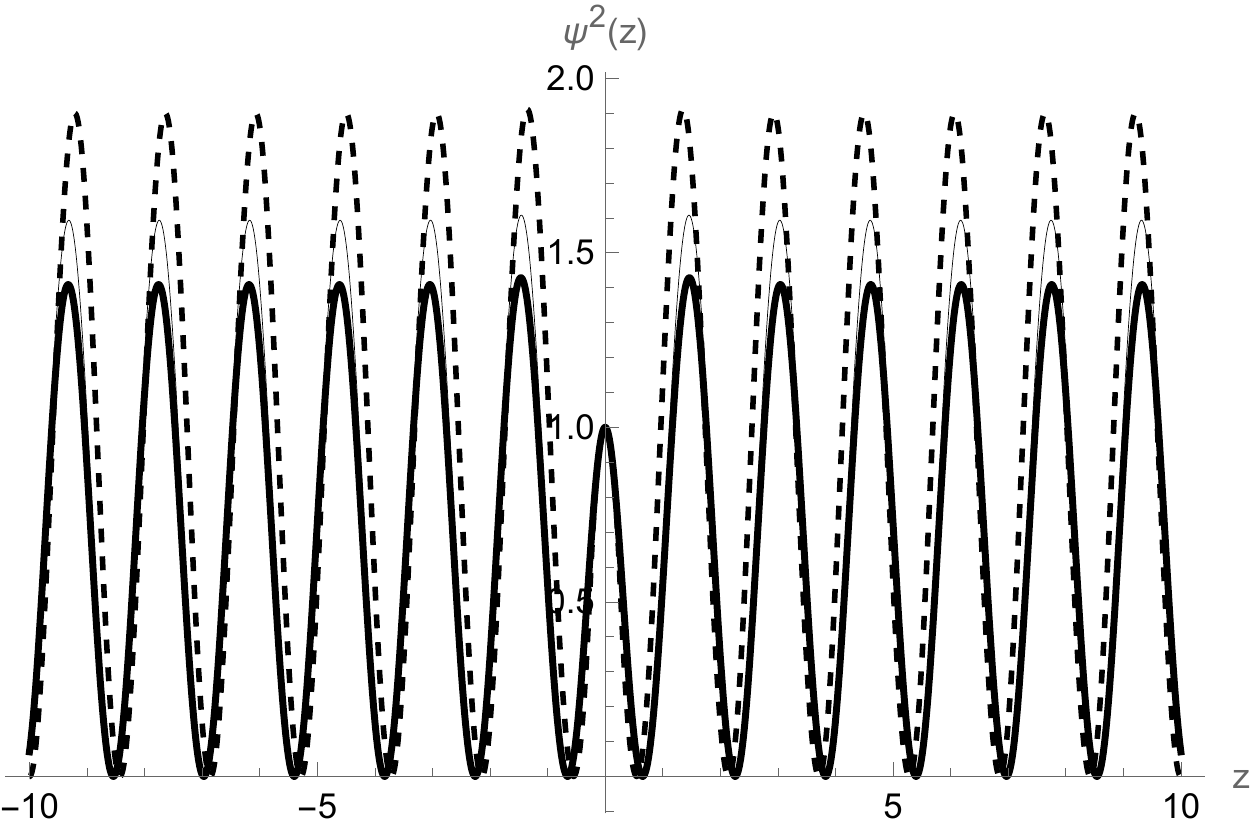}\\
           \caption{Left-handed massive mode for $m=4$. As $k$ increases from $k=0.005$ (thick line), $k=0.05$ (thin line) to $k=0.5$(dashed line) this mode keeps the same behaviour and the value $\psi^2 (0)$ is the same regardless of the value of $k$.}
           \label{nonquadraticmassive}
       \end{minipage}
\end{figure}

Finally, let us compare the effects of the variation of the $\sigma_i$ coefficients and of the $k$ on the massive modes. In fig.(\ref{quadraticmassive}) we plotted the squared wave-function $\psi^2$ for $m=4$. By varying the parameters, we see that the amplitude of the massive mode outside the brane and the decay rate of the mode inside the brane increase.

As long as $(\sigma_{1} + 2 \sigma_0)>0$, one obtains the usual $AdS_5$ warped 

\section{Final remarks} 
\label{sec5}
In this work we have studied a braneworld model where the gravitational dynamics is governed by a modified symmetric teleparallel gravitational theory. A non-minimal coupling between fermions and gravity, by means of the $\mathbb{Q}$ scalar, has also been investigated. 

By considering a quadratic gravitational Lagrangian $\mathbb{Q}$, the thin brane solutions, as in the RS model, can be embedded into a $dS_5$ or $AdS_5$ bulk, depending on the sign for the combination $\sigma_1 + 2\sigma_0= c_1+2c_5+8c_3$. We found that  $(\sigma_{1} + 2 \sigma_0)>0$ guarantees the existence of robust thin shell brane configurations in the $AdS_5$ bulk. For $\Lambda =0$, we also observed that the width of thick brane solutions is controlled by the effective parameter $\alpha=(\sigma_1+2\sigma_0)/6$. 

The modified non-quadratic Lagrangian $f(\mathbb{Q})=\mathbb{Q}+k\mathbb{Q}^2$ introduces new terms into the first-order BPS equations, leading to a 3-brane with inner structure that resembles a hybrid brane. Indeed, the potential wells of the sine-Gordon model become deeper, whereas the source scalar field tends to form a plateau at the origin. As a result, the brane undergoes a splitting process as the parameter $k$ is increased. These features are compatible with results found in other teleparallel braneworld models, as in $f(T)$ \cite{Yang2012,Menezes,allan1} and $f(T,B)$ \cite{allan} scenarios.

The non-minimal coupling between fermions and the nonmetricity scalar $\mathbb{Q}$ in the quadratic regime leads to results that are similar to those obtained with a standard Yukawa coupling. Thus, this geometric coupling can be interpreted as an alternative mechanism to get fermion localization. For $n=2$, the Schr\"{o}dinger-like potential undergoes a transition leading to a barrier and two symmetric wells around the origin. Nevertheless, the massless mode is still localized at the origin and the massive KK tower is still stable. This indicates that exotic couplings between matter and geometry may have mild effects on key aspects of matter localization in braneworld models.

Finally, the results presented here indicate that the different configurations studied are dependent on only two effective parameters $\sigma_0$ and $\sigma_1$, despite the fact that the scalar $\mathbb{Q}$ is constructed using up to $5$ different coefficients $c_i$. Moreover, key qualitative aspects of the resulting configurations are robust in large patches of the configuration space around the GR point ($\sigma_0=0$ and $\sigma_1=3/4$), which confirms the braneworld scenarios of GR as suitable phenomenological descriptions that transcend the limits of that theory. It is thus important to consider further investigations in dynamical scenarios and/or on non-flat $4D$ backgrounds, such as brane cosmology, stellar models, or black hole scenarios, which could help constrain the additional parameters of these models and put to a test the robustness of other predictions derived within the domain of GR. Research in this direction is currently under way.

\section*{Acknowledgments}
\hspace{0.5cm}The authors thank the Conselho Nacional de Desenvolvimento Cient\'{\i}fico e Tecnol\'{o}gico (CNPq), grants n$\textsuperscript{\underline{\scriptsize o}}$ 312356/2017-0 (JEGS), n$\textsuperscript{\underline{\scriptsize o}}$ 311732/2021-6 (RVM) and n$\textsuperscript{\underline{\scriptsize o}}$ 309553/2021-0 (CASA) for financial support. This work is supported by the Spanish Grants FIS2017-84440-C2-1-P and PID2020-116567GB-C21 funded by MCIN/AEI/10.13039/501100011033 (``ERDF A way of making Europe"), and the project PROMETEO/2020/079 (Generalitat Valenciana).


\begin{thebibliography}{10}

\bibitem{darkmatter}
C.~Boehm, P.~Fayet and R.~Schaeffer,
Phys. Lett. B \textbf{518} (2001), 8-14.

\bibitem{darkenergy}
A. G. Riess et al. 1998, AJ, 116, 1009.

\bibitem{Capozziello:2011et}
S.~Capozziello and M.~De Laurentis,
Phys. Rept. \textbf{509} (2011), 167-321
doi:10.1016/j.physrep.2011.09.003.


\bibitem{Hinterbichler:2011tt}
K.~Hinterbichler,
Rev. Mod. Phys. \textbf{84} (2012), 671-710
doi:10.1103/RevModPhys.84.671.



\bibitem{Maartens:2010ar}
R.~Maartens and K.~Koyama,
Living Rev. Rel. \textbf{13} (2010), 5
doi:10.12942/lrr-2010-5.



\bibitem{einsteincartan}
F.~W.~Hehl, P.~Von Der Heyde, G.~D.~Kerlick and J.~M.~Nester,
Rev. Mod. Phys. \textbf{48} (1976), 393-416.

\bibitem{metricaffine}
F.~W.~Hehl, J.~D.~McCrea, E.~W.~Mielke and Y.~Ne'eman,
Phys. Rept. \textbf{258}, 1-171 (1995).


\bibitem{fr}
A.~De Felice and S.~Tsujikawa,
Living Rev. Rel. \textbf{13} (2010), 3.

\bibitem{Arcos2004zh}
H.~I.~Arcos and J.~G.~Pereira,
Class. Quant. Grav. \textbf{21} (2004), 5193-5202
doi:10.1088/0264-9381/21/22/011.


\bibitem{fq3}
J.~Beltr\'an Jim\'enez, L.~Heisenberg and T.~Koivisto,
Phys. Rev. D \textbf{98}, no.4, 044048 (2018)





\bibitem{Aldrovandi} R. Aldrovandi and J. Pereira, Teleparallel Gravity. An
Introduction, Fundamental Theories of Physics Vol. 173
(Springer, Dordrecht, 2014).


\bibitem{Baez:2012bn}
J.~C.~Baez and D.~K.~Wise,
Commun. Math. Phys. \textbf{333} (2015) no.1, 153-186
doi:10.1007/s00220-014-2178-7.


\bibitem{Hohmann:2017duq}
M.~Hohmann, L.~J\"arv, M.~Kr\v{s}\v{s}\'ak and C.~Pfeifer,
Phys. Rev. D \textbf{97} (2018) no.10, 104042
doi:10.1103/PhysRevD.97.104042.


\bibitem{Maluf:2013gaa}
J.~W.~Maluf,
Annalen Phys. \textbf{525} (2013), 339-357
doi:10.1002/andp.201200272.



\bibitem{weit} R. Weitzenb\"{o}ck, Invarianten Theorie, Nordhoff, Groningen (1923).

\bibitem{ftinflation}
R.~Ferraro and F.~Fiorini,
Phys. Rev. D \textbf{75} (2007), 084031.

\bibitem{Cai}
Y.~F.~Cai, S.~Capozziello, M.~De Laurentis and E.~N.~Saridakis,
Rept. Prog. Phys. \textbf{79} (2016) no.10, 106901.

\bibitem{ftb}
S.~Bahamonde and S.~Capozziello,
Eur. Phys. J. C \textbf{77} (2017) no.2, 107.


\bibitem{fq1}
J.~Beltr\'an Jim\'enez, L.~Heisenberg and T.~S.~Koivisto,
JCAP \textbf{08}, 039 (2018)

\bibitem{fq2}
J.~Beltr\'an Jim\'enez, L.~Heisenberg, T.~S.~Koivisto and S.~Pekar,
Phys. Rev. D \textbf{101} (2020) no.10, 103507
J.~Beltr\'an Jim\'enez, L.~Heisenberg and T.~Koivisto,
Phys. Rev. D \textbf{98}, no.4, 044048 (2018)
\bibitem{delhom}
A.~Delhom,
Eur. Phys. J. C \textbf{80} (2020) no.8, 728.

\bibitem{fqenergyconditions}
S.~Mandal, P.~K.~Sahoo and J.~R.~L.~Santos,
Phys. Rev. D \textbf{102} (2020) no.2, 024057.


\bibitem{fqpropagation1}
M.~Hohmann, C.~Pfeifer, J.~Levi Said and U.~Ualikhanova,
Phys. Rev. D \textbf{99} (2019) no.2, 024009.

\bibitem{fqpropagation2}
I.~Soudi, G.~Farrugia, V.~Gakis, J.~Levi Said and E.~N.~Saridakis,
Phys. Rev. D \textbf{100} (2019) no.4, 044008.

\bibitem{fqcosmology1}
J.~Lu, X.~Zhao and G.~Chee,
Eur. Phys. J. C \textbf{79} (2019) no.6, 530.

\bibitem{fqcosmology2}
T.~Harko, T.~S.~Koivisto, F.~S.~N.~Lobo, G.~J.~Olmo and D.~Rubiera-Garcia,
Phys. Rev. D \textbf{98} (2018) no.8, 084043.

\bibitem{fqblackhole}
R.~H.~Lin and X.~H.~Zhai,
Phys. Rev. D \textbf{103} (2021) no.12, 124001

\bibitem{fqwormhole}
G.~Mustafa, Z.~Hassan, P.~H.~R.~S.~Moraes and P.~K.~Sahoo,
Phys. Lett. B \textbf{821} (2021), 136612.




\bibitem{rs1} 
  L.~Randall and R.~Sundrum,
  Phys.\ Rev.\ Lett.\  {\bf 83}, 3370 (1999).
 
\bibitem{rs2}

    L.~Randall and R.~Sundrum, Phys.\ Rev.\ Lett.\  {\bf 83} (1999), 4690. 


\bibitem{domainwall}
  O.~DeWolfe, D.~Z.~Freedman, S.~S.~Gubser and A.~Karch,
  Phys.\ Rev.\ D {\bf 62}, 046008 (2000).


\bibitem{Gremm}
M.~Gremm,
Phys. Lett. B \textbf{478}, 434-438 (2000).


\bibitem{wilami} W.~T.~Cruz, L.~J.~S.~Sousa, R.~V.~Maluf and C.~A.~S.~Almeida,
  Phys.\ Lett.\ B {\bf 730}, 314 (2014).


  
\bibitem{blochbrane} 
  D.~Bazeia and A.~R.~Gomes,
  JHEP {\bf 0405}, 012 (2004).
  
  
\bibitem{compacton}
D.~Bazeia, L.~Losano, M.~A.~Marques and R.~Menezes,
Phys. Lett. B \textbf{736} (2014), 515-521.

\bibitem{kehagias}
A.~Kehagias and K.~Tamvakis,
Phys. Lett. B \textbf{504} (2001), 38-46.



\bibitem{roldao} 
  J.~M.~Hoff da Silva and R.~da Rocha,
  Phys.\ Rev.\ D {\bf 81}, 024021 (2010).

\bibitem{minetic} 
  Y.~Zhong, Y.~P.~Zhang, W.~D.~Guo and Y.~X.~Liu,
  JHEP {\bf 1904}, 154 (2019), 	arXiv:1812.06453 [gr-qc].
 
 \bibitem{weyl} 
  Y.~X.~Liu, K.~Yang and Y.~Zhong,
  JHEP {\bf 1010}, 069 (2010).
  
\bibitem{conifold}     J.~E.~G.~Silva and C.~A.~S.~Almeida, Phys.\ Rev.\ D {\bf 84}, 085027 (2011).   

\bibitem{cigar}  
    J.~E.~G.~Silva, V.~Santos and C.~A.~S.~Almeida, Class.\ Quant.\ Grav.\  {\bf 30}, 025005 (2013).  
  
\bibitem{fR1}
D.~Bazeia, L.~Losano, R.~Menezes, G.~J.~Olmo and D.~Rubiera-Garcia,
Eur. Phys. J. C \textbf{75}, no.12, 569 (2015).

\bibitem{Yang2012}
J.~Yang, Y.~-L.~Li, Y.~Zhong and Y.~Li,
  Phys.\ Rev.\ D {\bf 85}, 084033 (2012).    

\bibitem{Menezes}
R.~Menezes,
Phys. Rev. D \textbf{89}, no.12, 125007 (2014).

\bibitem{ftnoncanonicalscalar}
J.~Wang, W.~D.~Guo, Z.~C.~Lin and Y.~X.~Liu,
Phys. Rev. D \textbf{98}, no.8, 084046 (2018).
  

\bibitem{allan1}
A.~R.~P.~Moreira, J.~E.~G.~Silva, F.~C.~E.~Lima and C.~A.~S.~Almeida,
Phys. Rev. D \textbf{103}, no.6, 064046 (2021).

\bibitem{Moreira:2021fva}
A.~R.~P.~Moreira, J.~E.~G.~Silva and C.~A.~S.~Almeida,
Int. J. Mod. Phys. D \textbf{30} (2021) no.13, 2150103. 
DOI:10.1142/S0218271821501030.


\bibitem{fqbrane} 
Q.~M.~Fu, L.~Zhao and Q.~Y.~Xie,
Eur. Phys. J. C \textbf{81} (2021) no.10, 890.



\bibitem{koley}
R.~Koley and S.~Kar,
Class. Quant. Grav. \textbf{22} (2005) no.4, 753-768.


\bibitem{fermionsinegordon}
Y.~X.~Liu, L.~D.~Zhang, L.~J.~Zhang and Y.~S.~Duan,
Phys. Rev. D \textbf{78} (2008), 065025.

\bibitem{casa}
 C.~A.~S.~Almeida, M.~M.~Ferreira, Jr., A.~R.~Gomes and R.~Casana,
  Phys.\ Rev.\ D {\bf 79}, 125022 (2009).
 
\bibitem{davi}
D.~M.~Dantas, D.~F.~S.~Veras, J.~E.~G.~Silva and C.~A.~S.~Almeida,
Phys. Rev. D \textbf{92} (2015) no.10, 104007.

\bibitem{allan}
A.~R.~P.~Moreira, J.~E.~G.~Silva and C.~A.~S.~Almeida,
Eur. Phys. J. C \textbf{81}, no.4, 298 (2021).

\bibitem{lyra}
J.~E.~G.~Silva, L.~J.~S.~Sousa, W.~T.~Cruz and C.~A.~S.~Almeida, Int. J. Mod. Phys. D {\bf 31} (2022) 2250030. DOI: 10.1142/S0218271822500304.

\end{thebibliography}
\end{document}